\documentclass[aps,reprint,prb,showpacs,showkeys,preprintnumbers,amsmath,amssymb,citeautoscript,floatfix,superscriptaddress]{revtex4-2}
\usepackage{graphicx}
\usepackage{pstricks}
\usepackage{hyperref}
\usepackage{subfigure}
\usepackage{dcolumn}
\usepackage{bm}

\usepackage{mathptmx}
\usepackage{import}
\usepackage[toc,page]{appendix}
\usepackage{chemformula}
\usepackage{amsmath}
\usepackage{ulem}
\DeclareMathOperator{\Tr}{Tr}

\newcounter{saveeqn}

\usepackage{placeins}

\begin{document}

\title{Machine learning assisted prediction of organic salt structure properties}

\author{Ethan P. Shapera}
\affiliation{Institute of Theoretical Physics and Computational Physics, Graz University of Technology, 8010 Graz, Austria}

\author{Dejan-Kre\v{s}imir Bu\v{c}ar}
\affiliation{Department of Chemistry, University College London, 20 Gordon Street, London WC1H 0AJ, United Kingdom}

\author{Rohit P. Prasankumar}
\affiliation{Enterprise Science Fund, Intellectual Ventures, 3150 139th Ave SE, Bellevue, WA 98005, United States}

\author{Christoph Heil}
\affiliation{Institute of Theoretical Physics and Computational Physics, Graz University of Technology, 8010 Graz, Austria}
\affiliation{Corresponding author: christoph.heil@tugraz.at}


\date{\today}

\begin{abstract}
We demonstrate a machine learning-based approach which predicts the properties of crystal structures following relaxation based on the unrelaxed structure.
Use of crystal graph singular values reduces the number of features required to describe a crystal by more than an order of magnitude compared to the full crystal graph representation.
We construct machine learning models using the crystal graph singular value representations in order to predict the volume, enthalpy per atom, and metal versus semiconducting phase of DFT-relaxed organic salt crystals based on randomly generated unrelaxed crystal structures. 
Initial base models are trained to relate 89,949 randomly generated structures of salts formed by varying ratios of 1,3,5-triazine and HCl with the corresponding volumes, enthalpies per atom, and phase of the DFT-relaxed structures. 
We further demonstrate that the base model is able to extrapolate to new chemical systems with the inclusion of 2,000 to 10,000 crystal structures from the new system.
After training a single model with a large number of data points, extension can be done at significantly lower cost.
The constructed machine learning models can be used to rapidly screen large sets of randomly generated organic salt crystal structures and efficiently downselect the structures most likely to be experimentally realizable.
The models can be used either as a stand-alone crystal structure predictor or incorporated into more sophisticated workflows as a filtering step.
\end{abstract}

\maketitle

\section{\label{sec:intro} Introduction}

Organic crystals form the basis of many common goods, including pharmaceuticals\cite{datta2004}, pesticides\cite{yang2017}, and pigments\cite{hao1997}, and have applications in emerging technologies such as thin film semiconductors\cite{kumar2014}, catalysts\cite{corma2010}, and optoelectronics\cite{bai2012}. 
An involved problem in materials design is the engineering of molecular crystals with targeted features (see e.g. Refs. \onlinecite{motherwell2002,oganov2011,price2014,reilly2016,corpinot2018}, among many others).
This task involves two major challenges: identifying a molecular target with promising solid-state properties, and the prediction and control of its crystal structure \cite{maddox1988,cruz2016,price2018,cheng2020}.
Crystal structure predictions (CSP) of organic solids are nowadays pursued using effective algorithms and great computational resources.
Yet, they have been shown to be very complex as unlike the constituents of inorganic crystals, organic molecules are generally conformationally flexible, resulting in numerous polymorphs with local energy minima within only a few kJ mol$^{-1}$ of the global minimum energy structure\cite{lommerse2000,price2014,nyman2015,greenwell2020}. 

Contemporary CSP approaches typically require 10,000's to 100,000's density functional theory (DFT) calculations to relax and calculate crystal energies and properties\cite{sontising2020,conway2021,nelson2021}, with the goal of identifying as many experimentally feasible crystal forms (e.g. polymorphs) as possible \cite{day2011,oganov2011,atahan2014,yang2018,curtis2018}.
The consideration of such large number of putative crystal structures causes, and the marginal differences in their energies render the identification of plausible structures exceedingly difficult. 
While approaches to lower the overprediction of organic crystal structures have recently been developed\cite{butler2023}, further steps need to be taken to make CSP more accurate, affordable and routine. 
With this in mind, we have developed a new machine learning approach to CSP\cite{oganov2011,atahan2014,yang2018,curtis2018} wherein we reduce the number of required energy calculations, thus lowering the computational cost of CSP, which otherwise can take up more than 100,000 CPU hours \cite{reilly2016}.

Machine learning is a rapidly developing tool for predicting properties of organic crystal structures that can address this issue.
The predictive power of machine learning models depends on the data used to train the model, the method for featurizing crystals, and the choice of machine learning algorithm.
Numerous studies have found success predicting properties of organic crystals, Refs. \onlinecite{musil2018,mcdonagh2019,egorova2020,wengert2021,balodis2022,kilgour2023}, by employing a wide range of machine learning approaches.
Common limitations are that either training data will contain crystals not in energetic local minima, which may not be representative of the relaxed structures, or generated structures need to be relaxed into an energetic minimum using ab initio methods, which again increases the computational cost.
The scope of organic CSP has been recently extended to organic solid solutions\cite{villeneuve2022}.

In this work, we demonstrate an approach to accelerate CSP by constructing machine learning (ML) models which can predict properties of DFT-relaxed organic crystals based on the structures and chemical compositions of randomly generated unrelaxed crystal structures.
By constructing ML models for enthalpy and volume of the corresponding relaxed structures, our approach allows downselection of structures which can then be explicitly evaluated using more expensive DFT calculations.
The downselection process removes randomly generated structures which are likely to relax into high energy configurations, leaving only the most relevant initial configurations.
Work with similar goals has been carried out by Honrao $\textit{et al.}$\cite{honrao2020} and Gibson $\textit{et al.}$\cite{gibson2022}.
Honrao $\textit{et. al}$ constructed support vector regression models for binary Al–Ni and Cd–Te systems featurized using radial and angular distribution functions.
Gibson $\textit{et. al.}$ trained crystal graph convolutional neural networks to predict formation energies using inorganic structures from the Materials Project Database.

\begin{figure}
\includegraphics[width=\linewidth]{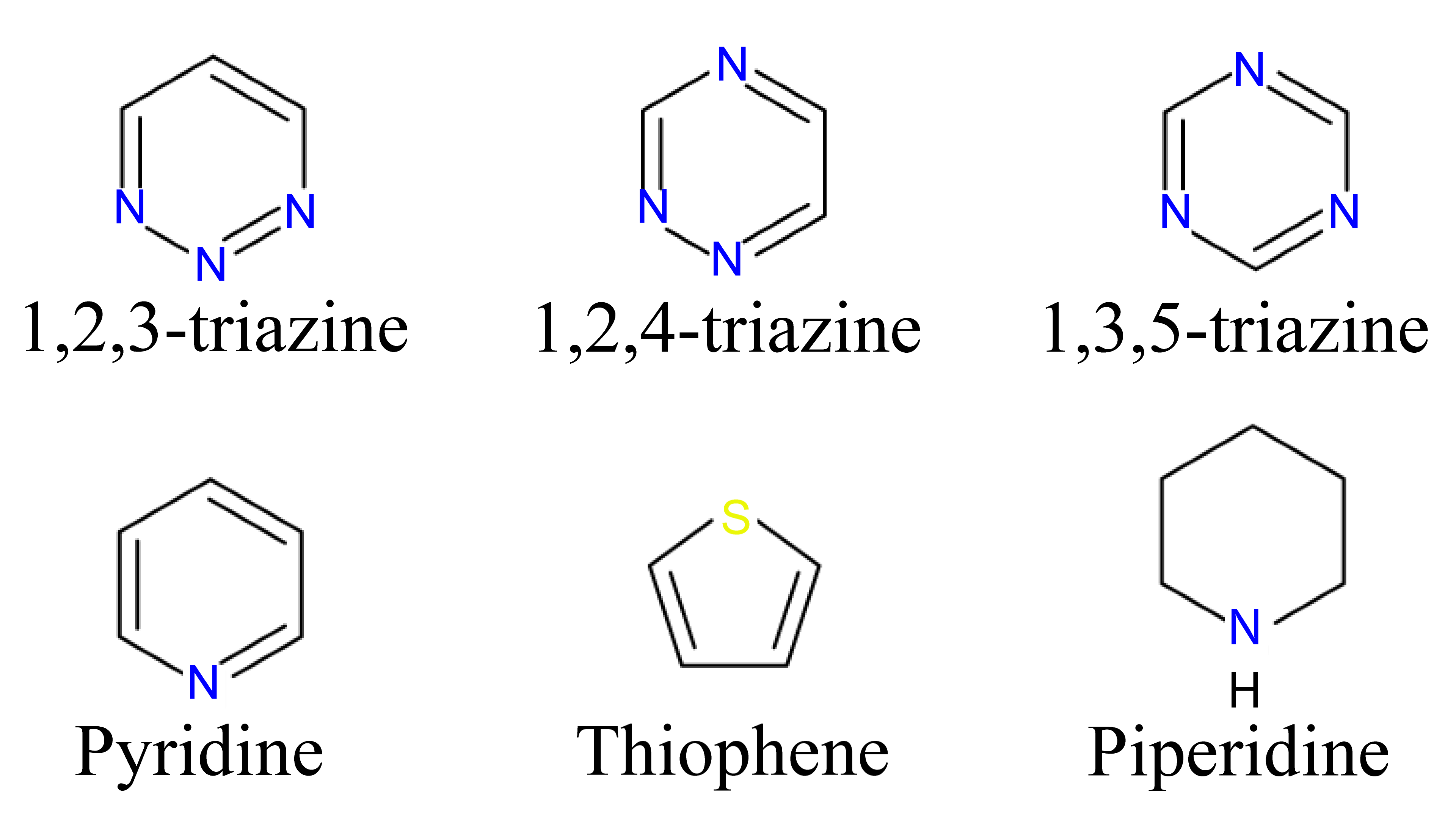}
\caption{\label{fig:Molecules}
Diagrams of molecules used in this study. From left to right: 1,2,3-triazine, 1,2,4-triazine, 1,3,5-triazine, pyridine, thiophene, and piperidine.}
\end{figure}

Presented here are two further developments on machine learning modeling for CSP.
First, we describe and validate a crystal graph singular value representation of crystal structures which reduces the required number of descriptors by several orders of magnitude.
Crystal graph descriptors were developed by Xie \textit{et al.}\cite{xie2018} as an effective method for featurizing crystal structures which can be used to train high accuracy convolutional neural network (CGCNN) models.
The approach has been used in numerous studies, including Refs. \onlinecite{chen2019,park2020,karamad2020,lee2021,gibson2022}, yet requires a large number of values to describe each structure, causing data files to become quite large. 
As an example, in the case of 11-atom C$_3$H$_3$N$_3$ HCl, each atom is represented by a $12\times41$ matrix, for a total of 5,412 descriptors required to represent the crystal structure.
By using the singular values, the representation is reduced to fewer than 300 values.
Second, we employ random forest models which require fewer hyperparameters and are fit without the iterative backpropagation process required for neural network algorithms\cite{breiman2001}. 
The machine learning approach developed in this work provides a method for predicting properties of relaxed organic crystals knowing only the structure of the randomly generated unrelaxed structures while requiring only a small number of DFT relaxations to form a training set.
We apply our machine learning methods to salts formed from HCl (and HBr in Extension A) and six small organic ring molecules: 1,2,3-triazine, 1,2,4-triazine, 1,3,5-triazine, pyridine, thiophene, and piperidine, shown in Figure\ \ref{fig:Molecules}.
The choice of acid to protonate the organic molecule and the relative concentrations of acid and organic molecule influence the crystal structure of the targeted solid.

The machine learning approach we develop here does not make any assumptions about the chemical composition of the system, the range of structures, the method used to generate trial structures, or the approach used to optimize structures after generation.
This makes the approach broadly applicable to both organic and inorganic crystals.
Further, our approach may be incorporated into more sophisticated CSP workflows.
Our work will demonstrate a machine learning approach which is broadly applicable to organic crystals and may be incorporated into sophisticated CSP workflows as a filter to remove thermodynamically unfavorable structures.
We especially encourage attempts to integrate our approach into organic CSP based on machine learning-trained interatomic potentials.

\section{\label{sec:methods} Methods}

\subsection{\label{subsec:overview}Overview}
Complete descriptions of the choices of molecular crystals, DFT and CSP methods, model construction, and model evaluation are included in the Supplemental Materials sections \ref{subsec:MoleculeChoice} - \ref{subsec:modelEvaluate}.

Random crystal structures were generated using the \texttt{AIRSS} \cite{pickard2006,pickard2011} software package.
Organic molecules were initially generated with fixed 2D structures.
During relaxation, the unit cell shape, relative positions and orientations between molecules, and geometries of individual molecules were optimized simultaneously.

Base models are constructed to relate descriptors of unrelaxed structures to DFT-computed unit cell volume ($V$), enthalpy per atom ($h$) and metal or semiconductor phase (phase).
The base ML models are fit using data from four crystal structure prediction runs for 1,3,5-triazine.
In order to test the ability of the learned model to predict properties of crystals of different organic molecule salts, we consider three levels of extension,
\begin{itemize}
    \item Extension A: 1,3,5-triazine HBr, 1,2,3-triazine HCl, and 1,2,4-triazine HCl
    \item Extension B: pyridine (C$_5$H$_5$N) HCl
    \item Extension C: thiophene (C$_4$H$_4$S) HCl and piperidine (C$_5$H$_{11}$N) HCl
\end{itemize}
Extension A includes two sets of structures in which the organic molecule is an isomer of 1,3,5-triazine and an organic salt of 1,3,5-triazine with HBr instead of HCl.
Extension B considers salts of pyridine, another small nucleophillic molecule based on a six member ring.
CSP for pyridine has been previously shown to be a challenging problem due to having numerous computed structures in the energy gap between the observed and most stable structure as well as a complex, asymmetric unit cell \cite{anghel2002}.
Extension C contains thiophene HCl and piperidine HCl.
Extension C is the largest extension away from 1,3,5-triazine HCl with thiophene consisting of a five member heterocyclic molecule and piperidine an example of a less nucleophilic molecule.
We also note that the thiophene sulfur atom does not protonate easily and therefore there is a lack of known structures of protonated thiophene derivatives, but is included to test model performance on high energy structures.

Constructed models are tested for overfitting using a nested cross validation scheme.
The accuracy of each regressor model is quantified using the mean absolute error (MAE), mean absolute fractional error (MAFE), and Spearman coefficient ($\rho$). 
Classifier models are evaluated using the average precisions (AP) for each class and mean average precision (mAP) based on the precision-recall curves.

\subsection{\label{subsec:descriptors} Descriptor Choice}
For every crystal structure, three sets of descriptors are compiled: crystal graph singular values, Coulomb matrix eigenvalues, and crystal structure parameters.

Crystal graph representations of every structure are generated using code provided by Xie \textit{et al.} \cite{xie2018} on the authors' Github repository\cite{xieGit}.
The approach describes local chemical environments for each atom as graphs characterizing the bonding arrangements between atoms.
An $n \times m$ matrix, $c_i$, is constructed to represent each of the $a$ atoms in the unit cell.
The crystal graph matrix for each atom in the unit cell is used to construct an $an \times am$ block diagonal matrix $B$, 
\begin{equation}
    \label{eq:block}
    B = c_1 \oplus c_2 \oplus ... \oplus c_a.
\end{equation}
The singular values of $B$ are used as descriptors for our ML models, and are referred to as Crystal Graph Singular Value (CGSV) descriptors.
Performance of models including CGSV descriptors is discussed in \ref{subsec:includeCG}.
The code by Xie \textit{et al.} was used solely to generate the crystal graph representations; no results were generated from the authors' pretrained models.

The Coulomb matrix $C_{ij}$ of a structure is defined as:
\[
    C_{ij}= 
\begin{cases}
    \frac{Z_iZ_j}{r_{ij}},& \text{if } i\neq j\\
    \frac{1}{2}Z_i^{2.4},              & \text{if } i=j
\end{cases}
\]
where $Z_i$ is the atomic number of atom $i$ and $r_{ij}$ is the distance between atoms $i$ and $j$ taking into account periodic boundary conditions\cite{rupp2012,montavon2012}.
The Coulomb matrix descriptors are then found as the sorted eigenvalues of $C_{ij}$.
Further descriptors are formed from: number of positive eigenvalues, number of negative eigenvalues, $\Tr(C_{ij})$, and $\det(C_{ij})$.

We note that $\Tr(C_{ij}) = \sum_k \lambda_k = \frac{1}{2}\sum_k Z_k^{2.4}$ where $\{\lambda_i\}$ are the eigenvalues of the Coulomb matrix.
The trace of the Coulomb matrix, then, provides a unique irrational number which characterizes the atomic composition of the unit cell.
This allows the trace to be used as a categorical descriptor to identify the atomic contents of the cell.
However, this trace descriptor only identifies the chemical contents of the cell, not the bonding arrangement.
The descriptor formed from $\Tr(C_{ij})$ distinguishes, e.g. 1,3,5-triazine (C$_3$H$_3$N$_3$) from pyridine (C$_5$H$_5$N), but would not distinguish 1,3,5-triazine from 1,2,3-triazine.

The crystal structure descriptors consist of the unit cell edge lengths A,B, and C and unit cell angles $\alpha$, $\beta$, $\gamma$. 

While all molecules studied here are rigid and flat, the chosen descriptors are applicable to more complex 3D and conformationally flexible molecules.

\subsection{\label{subsec:modelExtrapolate} Model Extension}

The ability of models to extrapolate to crystals with different chemical compositions is tested by incrementally incorporating structures from an extension CSP run into the 1,3,5-triazine HCl training sets.
Structures in four 1,3,5-triazine HCl CSP runs are randomly divided into fitting, validation, and testing sets as described in \ref{subsec:modelConstruct}.
$N$ structures from the extension CSP run are randomly added to the fitting set and all remaining extension CSP structures are added to the testing set.
The fitting set is weighted such that the total weight of 1,3,5-triazine HCl structures is four, corresponding to four CSP runs, and the total weight of the added CSP run structures in the fitting set is one.
Random forest models are fit using the combined fitting set then applied to the validation and testing sets.
Performance of the random forest models are evaluated for the 1,3,5-triazine HCl runs and extension run separately.
Monitoring separate performances for 1,3,5-triazine HCl and the included extension run checks decreased accuracy for 1,3,5-triazine HCl due to inclusion of new structures and ability of the model to extrapolate outside the 1,3,5-triazine HCl data set.

\section{\label{sec:results}Results and Discussion}

\subsection{\label{subsec:descriptorSelect} Descriptor Selection}

\begin{figure}
\includegraphics[width=\linewidth]{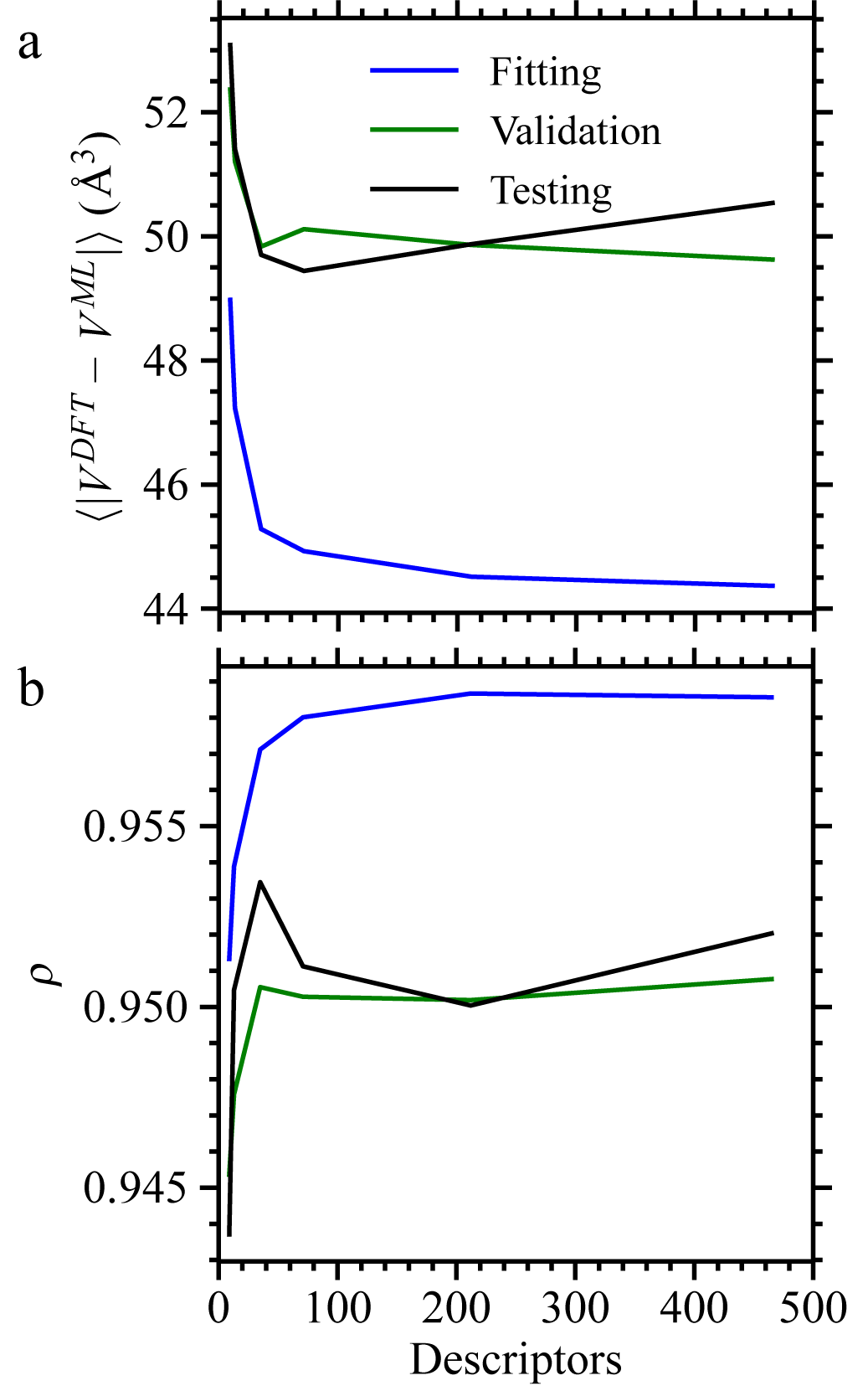}
\caption{\label{fig:NumberDescriptors}
 (a) MAE for volume model trained on 1,3,5-triazine HCl crystal structures versus number of descriptors included in model construction.
 (b) Spearman correlation coefficient for volume model trained on 1,3,5-triazine HCl crystal structures versus number of descriptors included in model construction.
}
\end{figure}

We minimize the random forest regressor model training errors by optimizing the number of descriptors to be used for model construction.
An initial decision tree regressor is fit to the full set of 506 descriptors with a maximum tree depth of 20 layers with the model Gini importances for all descriptors tabulated.
The Gini importance of a descriptor quantifies the importance of a descriptor in a tree-based model by considering both the number of times each descriptor is used in the fitted model and the number of samples split by the descriptor \cite{breiman2001,menze2009,scikit-learn}.
Features with Gini importances greater than "$c_i \times$ \textit{average Gini importance}" are retained for fitting subsequent random forest regressors.
$c_i$ is assigned values between 0.01 and 1.0, with the corresponding number of descriptors dependent on the choices of fitting set and target quantity.
Figure\ \ref{fig:NumberDescriptors}(a) shows the corresponding fitting, validation, and testing mean absolute errors (MAEs) for the constructed random forest regressors.
The Spearman correlation coefficients between $V^{DFT}$ and $V^{ML}$ are plotted against the number of included descriptors in Figure\ \ref{fig:NumberDescriptors}(b).
The fitting MAE decreases monotonically with increasing number of descriptors, which we attribute to the increasing number of possible splitting criteria for fitting random forest regressors.
The testing set MAE shows a minimum at 70 included descriptors.
For the number of descriptors below 70, the number of splitting criteria is not sufficient to capture the relation between the descriptors and $V$.
However, increasing the number of descriptors to fit the random forest models beyond 70 leads to more severe overfitting, which manifests itself in the decreasing fitting set MAE, while the testing MAE increases.
With more than 13 descriptors included, the Spearman coefficients for the fitting, validation, and testing sets are all above 0.95.
Similar results are obtained for models constructed for the \textit{enthalpy per atom} regressor and \textit{metal versus semiconductor} classifier.
From these results, we determine an appropriate criterion for downselecting descriptors: include only descriptors with Gini importances greater than 0.1 times the average Gini importance.

The number of descriptors selected for each model in this paper are listed in Table \ \ref{table:NumDescriptors}.
The number of descriptors used for each target quantity shows little change based on the CSP run added to the fitting set.
For example, in constructing models for $V$, the number of descriptors included varies between 63 and 70.
The only large difference is observed in the model of $h$ for thiophene HCl, which takes 56 descriptors.
For the other five $h$ models, 110 to 122 descriptors are selected.

\begin{figure}
\includegraphics[width=\linewidth]{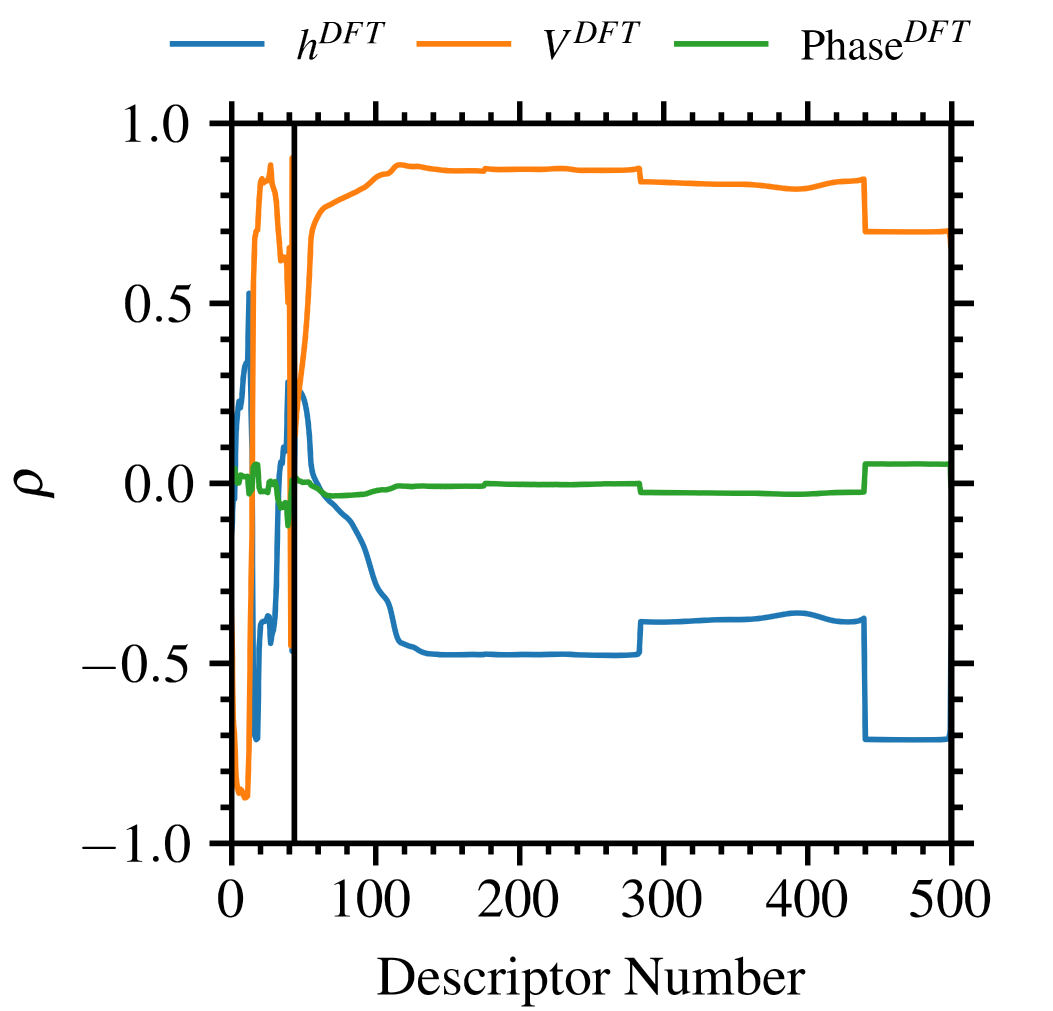}
\caption{\label{fig:SpearmanCorr}
 Spearman correlation coefficients between target quantities and Coulomb matrix and crystal graph singular value descriptors for the four datasets of 1,3,5-triazine HCl CSP.
 The orange curve plots $\rho$ for $V^{DFT}$, blue curve plots $\rho$ for $h^{DFT}$ and the green curve plots $\rho$ for metal versus insulator from DFT.
 Descriptors 0 through 44 to the left of the vertical black line corresponds to Coulomb matrix descriptors while descriptors 45 through 500 correspond to the crystal graph singular value descriptors.
}
\end{figure}

The required number of descriptors can be rationalized by consideration of the Spearman correlation coefficients ($\rho$) between the descriptors and target quantities, shown in Figure \ \ref{fig:SpearmanCorr}.
The strongest correlations between descriptors and target values are found for $V^{DFT}$.
Out of 506 possible descriptors, 380 descriptors have $|\rho| \geq 0.8$ with $V^{DFT}$ and only 6 descriptors have $|\rho| \leq 0.2$.
This indicates that the chosen descriptors are strongly correlated with $V^{DFT}$.
For $h$ and phase, no descriptors with $|\rho| \geq 0.8$ are present.
Strong correlations between descriptor and target values allows the construction of low error models with fewer fitting parameters.
ML models constructed for $V^{DFT}$ would be expected to require fewer descriptors than for $h^{DFT}$ and phase.
Of the three target quantities, $V$ has the most direct connection to the chosen descriptors.
To take the Coulomb matrix as an example, the off-diagonal elements have a $1/r_{ij}$ dependence.
Increasing the volume of the unit cell will increase the typical distances between atoms in the cell, thereby decreasing the off-diagonal elements of the Coulomb matrix and the magnitudes of the eigenvalues.
Similarly, the off-diagonal elements of the Coulomb matrix are the potential energies due to the Coulomb interactions between pairs of nuclei, thereby providing a measure of one source of potential energy in the crystal.
However, the Coulomb matrix neglects all other contributions to the enthalpy, such as electron-electron and electron-nucleus interactions.
There is no direct connection between the chosen descriptors and the phase.

\begin{table*}[t]
\begin{tabular}{c c c c}
\hline
Added CSP Set & \begin{tabular}{@{}c@{}}Number of Descriptors \\ Target Quantity - $V$\end{tabular} & \begin{tabular}{@{}c@{}}Number of Descriptors \\ Target Quantity - $h$\end{tabular} & \begin{tabular}{@{}c@{}}Number of Descriptors \\ Target Quantity - Phase \end{tabular} \\ \hline
None & 70 & 122 & 265\\
1,3,5-triazine HBr & 63 & 116 & 268\\
1,2,3-triazine HCl & 64 & 109 & 265\\
1,2,4-triazine HCl & 65 & 116 & 274\\
Pyridine HCl & 67 & 110 & 260\\
Thiophene HCl & 63 & 56 & 265\\
Piperidine HCl & 60 & 45 & 252\\
\hline
\end{tabular}
\caption{\label{table:NumDescriptors}
Number of Descriptors used for each model.
"Added CSP set" of "None" indicates that only the four CSP runs with 1,3,5-triazine HCl were included.
For entries with a listed dataset under "Added CSP set", 10,000 structures from the added set were included in along with 80\% of the structures from the four 1,3,5-triazine CSP runs in fitting the decision tree regressor.
}
\end{table*}

\subsection{\label{subsec:triazineModel} 1,3,5-triazine Crystal Base Model Construction}

All results in this subsection refer to models trained with the four datasets of 1,3,5-triazine with HCl listed in Table \ref{table:molecules}.

\begin{figure}
\includegraphics[width=\linewidth]{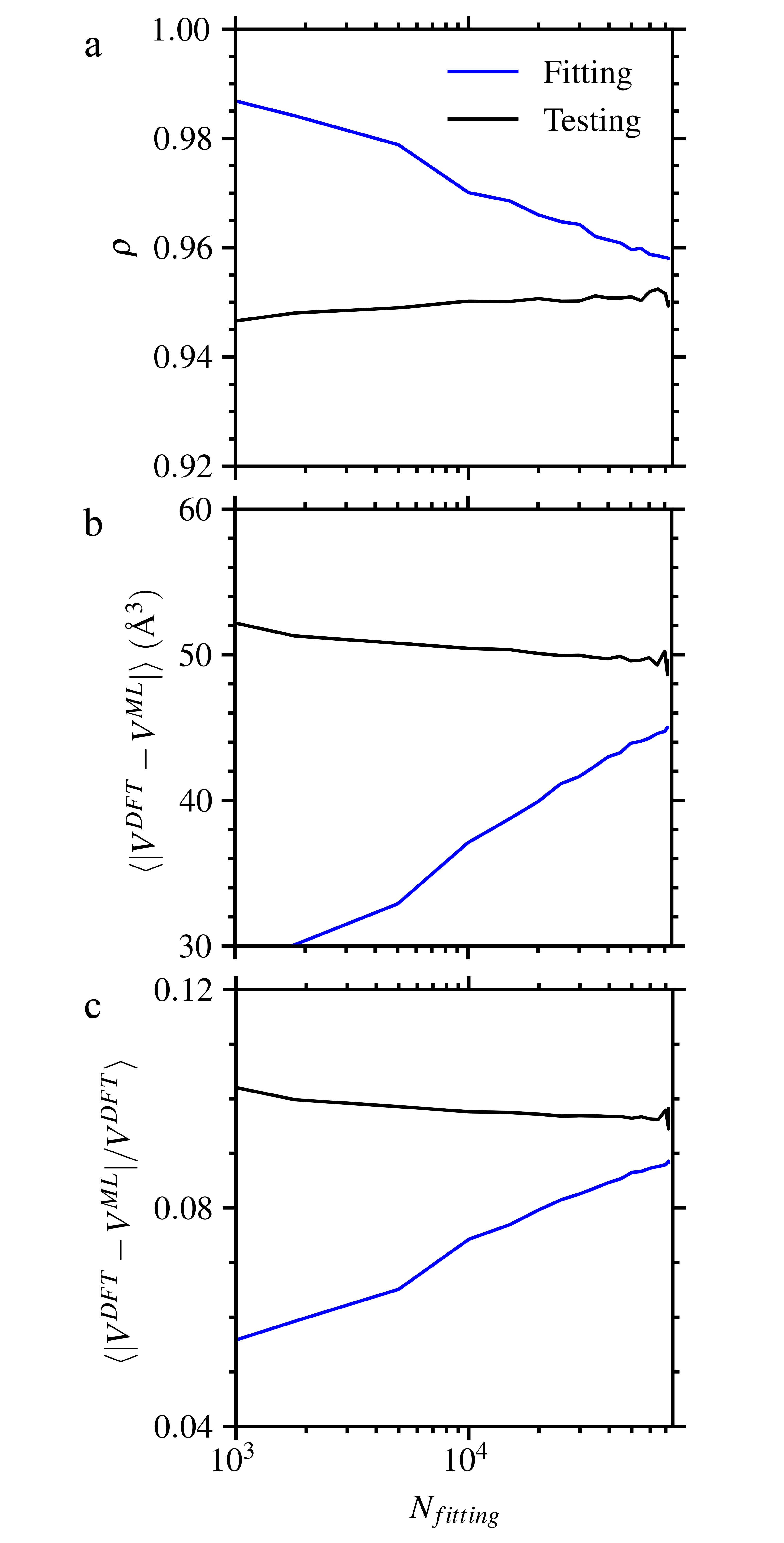}
\caption{\label{fig:TriazineFittingSize}
Performance of the base 1,3,5-triazine HCl $V$ model versus number of randomly selected samples in the fitting set, characterized by (a) Spearman coefficient, (b) MAE, and (c) MAFE.
}
\end{figure}

\textbf{Fitting set size.}
A persistent consideration in constructing ML models is the number of datapoints required to train the model.
If too few datapoints are used in training, the ML algorithm may be unable to find patterns relating the descriptors and target values, resulting in either overfit or underfit models.
Using arbitrarily large training sets is also undesirable because it increases the cost to generate the training set and the cost to fit the model.
As an example, we consider performance of random forest regressor models to predict crystal structure volumes which are trained with varying fitting set sizes.
Results are plotted in Figure\ \ref{fig:TriazineFittingSize}.
From the four CSP runs with 1,3,5-triazine and HCl, $9 \times N_{fitting}/8$ structures are randomly selected to form the fitting and validation sets.
All remaining structures are placed in the testing set.
As in the rest of this work, we iterate model construction over 10 random splittings of the fitting and validation sets with $N_{fitting}$ structures in each fitting set and $N_{fitting}/8$ structures in each validation set.
For the testing set, values of the MAE, MAFE, and Spearman coefficient are consistent over the range $5,000 \leq N_{fitting} \leq 70,000$.
Improvement in the model with respect to $N_{fitting}$ is observed through the increase in MAE and MAFE of the fitting set and decrease in the difference of the Spearman coefficients for the fitting set and testing set.
Adding more fitting data increases the model error for the fitting data, while decreasing the overfitting of the model.
Moving forward, we use the full data set of 1,3,5-triazine and HCl for the base model.
Due to our validation scheme, this corresponds to 71,959 structures in the fitting set for each iteration.

\begin{figure}
\includegraphics[width=0.8\linewidth]{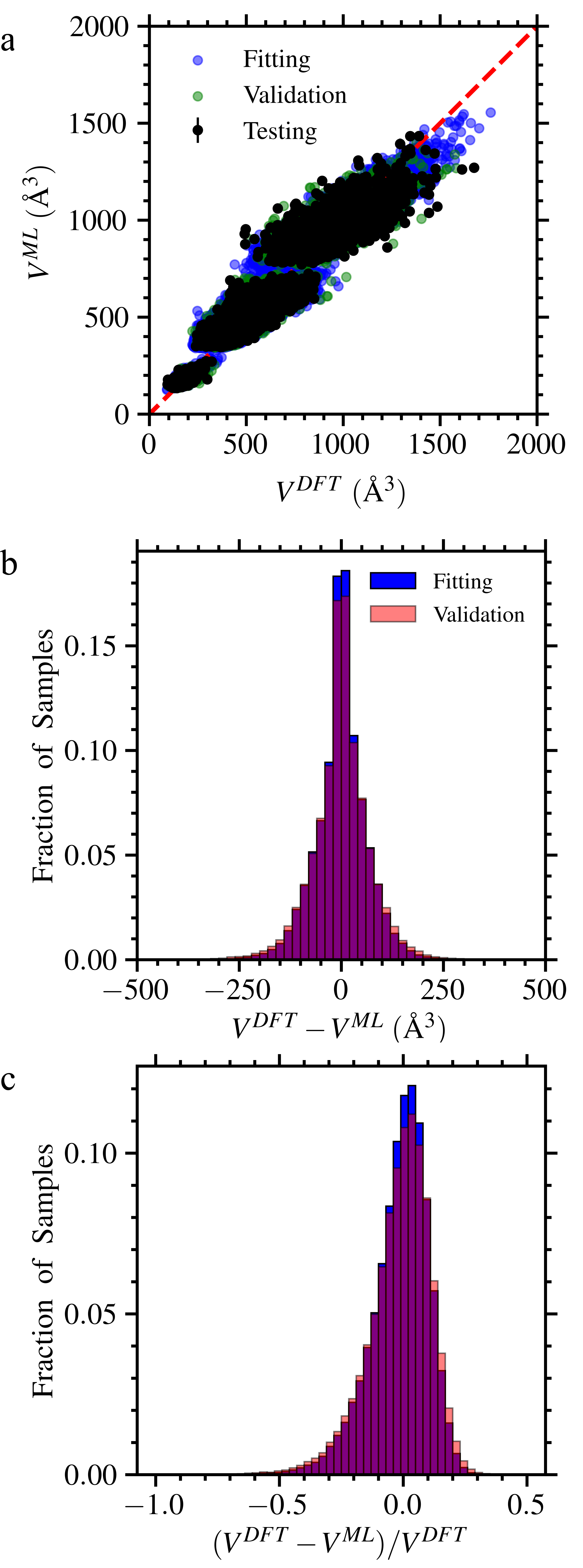}
\caption{\label{fig:TriazineModel}
  (a) Parity plot comparison of $V^{DFT}$ obtained via DFT and $V^{ML}$ predicted by random forest regressor. Plotted are the results from 1 fitting-validation iteration and testing set average values and error bars from all 10 iterations. Dashed red line guides the eye to show $V^{DFT}=V^{ML}$.
  (b) Distribution of ML errors of $V$ for 10 fitting-validation iterations. 
  (c) Distribution of ML fractional errors in $V$ for 10 fitting-validation iterations.
}
\end{figure}

\textbf{Volume model.} 
One design guideline for determining which crystal polymorphs are experimentally realizable is that the structure should have high density, with a packing density of 60\% to 80\% \cite{kitaigorodskii1961,corpinot2018}.
Preferring high density crystal structures removes highly porous structures with large voids.
Porous structures are often unfavorable due to the ability of constituent atoms and molecules to rearrange into lower energy configurations by filling the voids.
Random forest regressor trained on DFT calculated volumes and comparison of the machine-learned and DFT-calculated volumes for the 1,3,5-triazine HCl training set is shown in Figure \ref{fig:TriazineModel}.
The model produces MAEs of 45 $\AA^3$ for the fitting set, 50 $\AA^3$ for the validation set, and 49 $\AA^3$ for the testing set.
This indicates that with the optimum choice of the maximum tree depth hyperparameter, overfitting can be minimized, but is still observable:
the validation and testing MAEs are approximately factors of 1.1 and 1.09 larger than the fitting MAE, respectively.
While the validation and testing MAEs are larger than the fitting MAE, the discrepancies are small.

The random forest regressor model produces non-Gaussian distributions of fitting and validation errors in $V^{\mathrm{ML}}$,
see Figure\ \ref{fig:TriazineModel}(b).
This plot shows the differences between machine-learning predicted $V^{\mathrm{ML}}$ and DFT-calculated $V^{\mathrm{DFT}}$ for fitting and validation sets over all 10 fitting-validation iterations.
The fitting set shows a mean error of 0.018 $\AA^3$ with 1-standard deviation width of 62 $\AA^3$.
The validation set has a mean error of -0.10 $\AA^3$ with 1-standard deviation width of 70 $\AA^3$.
We observe a small bias toward underestimating $V^{\mathrm{DFT}}$, with the constructed model predicting smaller values for 51\% of both fitting and validation materials.
The unimodal distribution of errors indicates that there is no subgroup of initial structures for which the model consistently fails.
Further, the Spearman correlation coefficients between the DFT-calculated volumes and the model predicted volumes are calculated as 0.95 for both the fitting and validation sets.
By considering the fractional error distributions in Figure\ \ref{fig:TriazineModel}(c), bias and non-Gaussian behavior in the constructed model is further shown.
The fitting set shows a mean fractional error of -1.4\,\% and the validation set has a mean fractional error of -1.6\,\%.
While the absolute errors show nearly symmetrical distributions, the fractional errors are skewed, with skewnesses of -0.97 for fitting and -1.01 for validation, which manifests as a tail in the fractional error distribution seen for $V^{\mathrm{DFT}} < V^{\mathrm{ML}}$.

\begin{figure}
\includegraphics[width=\linewidth]{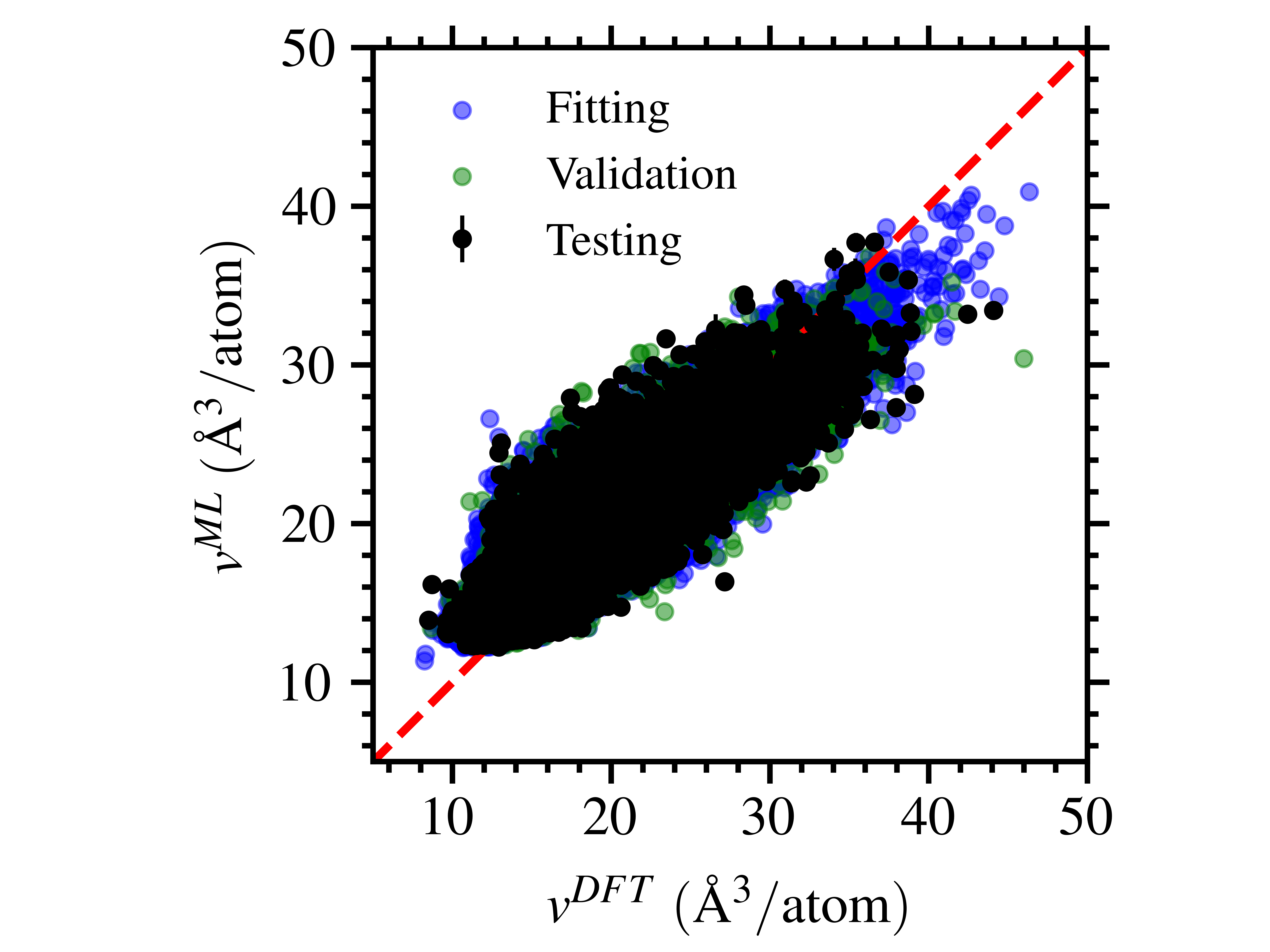}
\caption{\label{fig:TriazineVolPerAt}
  Parity plot showing the performance of the volume model.
  Data shown corresponds to Figure \ref{fig:TriazineModel} (a) with volumes normalized per atom in the unit cell.}
  Plotted are the results from 1 fitting-validation iteration and testing set average values and error bars from all 10 iterations
  Dashed red line guides the eye to show $v^{DFT}=v^{ML}$.
\end{figure}

Volume is an extensive variable which depends strongly on the number of atoms in the unit cell.
One consequence is clusters visible in Figure\ \ref{fig:TriazineModel} (a) corresponding to different CSP runs with different numbers of atoms in the cell.
If in a CSP run multiple different values for the number of organic molecule units in the cell are included, the volume minimization criterion would select structures with the fewest organic molecules.
Taking into account different cell contents requires using the intensive quantity of volume per atom $v$.
Normalizing the volumes plotted in Figure\ \ref{fig:TriazineModel} (a) per atom in each CSP run produces the parity plot in Figure\ \ref{fig:TriazineVolPerAt}.
The model trained on total volume retains high Spearman correlation coefficients of 0.96 for the fitting set, 0.95 for the validation set, and 0.88 for the testing set when applied to volume per atom.
The model trained on total volume is readily adapted to predict volume per atom.
This provides the potential to train on small unit cells and extrapolate to larger cells.

For this work, we did not find it necessary to spend excessive resources developing highly accurate ML models of the volume of relaxed crystal structures.
However, the model still outperforms selecting generated structures based solely on the unrelaxed volume, as shown in a test case in the Supplemental Materials section \ref{subsec:baseline}.
Instead we use a less accurate model which reproduces general trends as a coarse filter and can then explicitly check the predictions by performing DFT simulations for downselected randomly generated initial unrelaxed structures.
While the volume model does not produce sufficiently accurate predictions to replace explicit DFT calculation, it does show remarkable agreement in error distributions between the fitting, validation, and testing sets.

\begin{figure}
\includegraphics[width=0.8\linewidth]{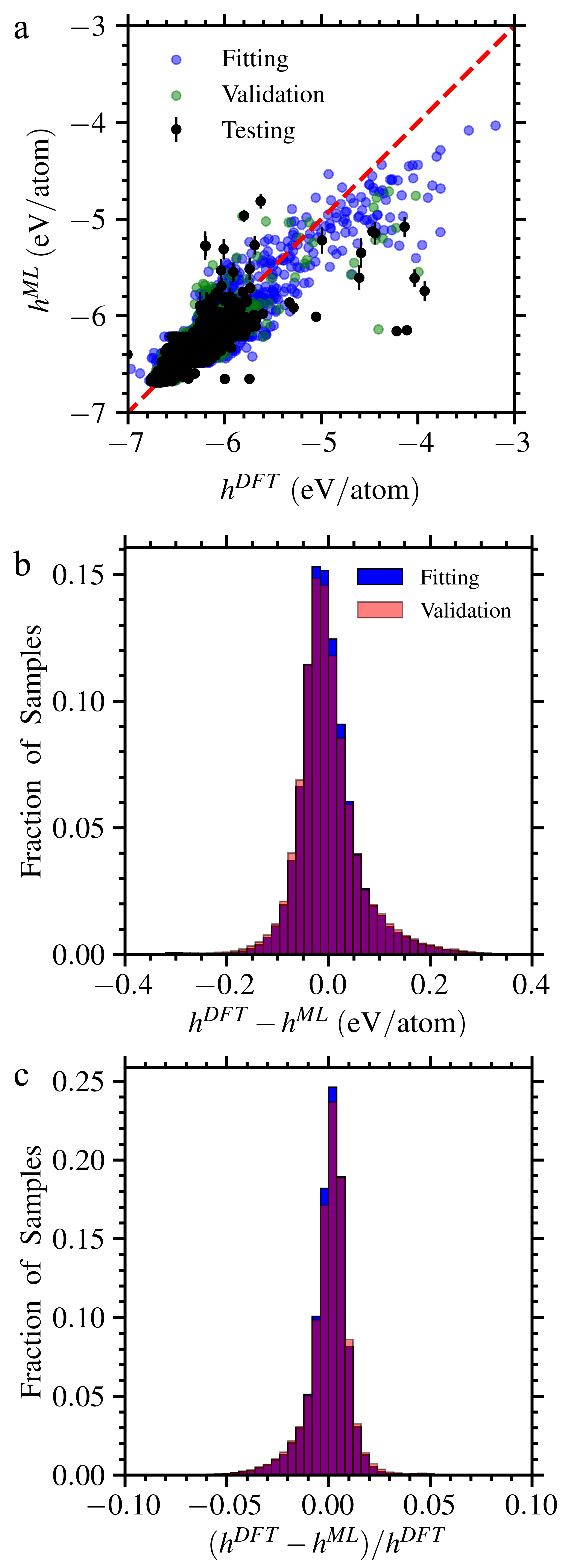}
\caption{\label{fig:TriazineModelEnthalpy}
  (a) Parity plot comparison of $h^{DFT}$ obtained via DFT and $h^{ML}$ predicted by random forest regressor. Plotted are the results from 10 fitting-validation iterations as well as testing set average values and error bars from all 10 iterations.
  (b) Distribution of ML errors of $V$ for 10 fitting-validation iterations. 
  The distributions are unimodal and symmetric about their means.
  (c) Distribution of ML fractional errors in $V$ for 10 fitting-validation iterations.
}
\end{figure}

\textbf{Enthalpy model.} 
Our second criterion for selecting polymorphs is the enthalpy per atom of the structure.
Enthalpy is chosen as a thermodynamic criterion because it includes consideration of pressure and volume effects in the CSP. 
A random forest regressor trained on DFT calculated enthalpies per atom and comparison of the machine-learned and DFT-calculated enthalpies per atom for the 1,3,5-triazine HCl training set is shown in Figure \ref{fig:TriazineModelEnthalpy}.
The model produces MAEs of 0.044 eV/atom for the fitting set, 0.048 eV/atom for the validation set, and 0.047 eV/atom for the testing set. 
This indicates that with the optimum choice of the maximum tree depth hyperparameter, overfitting is still present, but minimal:
the validation and testing MAEs are approximately factors of 1.09 and 1.07 larger than the fitting MAE, respectively.
As in the volume model, the MAE values for the validation and testing sets closely match the MAE of the fitting set.

The model for $h$ has MAFE values of 0.0070 for the fitting set, 0.0077 for the validation set, and 0.0076 for the testing set.
The enthalpy per atom model produces MAFE values which are an order of magnitude smaller than the MAFE values of the volume model. 

For the purpose of downselecting structures to relax, one would not be interested in crystal structures over the entire range of $h^{ML}$, only low $h^{ML}$ structures. 
Considering only the testing set structures with the 1,000 lowest $h^{ML}$ values, the MAE drops to 0.026 eV/atom and the MAFE to 0.0040.
Thus, the $h$ model is considerably more accurate in the region of interest.

The random forest regressor model produces non-Gaussian distributions of fitting and validation errors in $h^{\mathrm{ML}}$,
see Figure\ \ref{fig:TriazineModelEnthalpy}(b).
This plot shows the differences between machine-learning predicted $h^{\mathrm{ML}}$ and DFT-calculated $h^{\mathrm{DFT}}$ for fitting and validation sets over all 10 fitting-validation iterations.
The fitting set shows a mean error of $1.0\times 10^{-4}$ eV with 1-standard deviation width of 0.066 eV.
The validation set has a mean error of $1.6\times 10^{-5}$ eV with 1-standard deviation width of 0.080 eV.
Errors between the DFT-computed and the ML predicted enthalpies per atom show unimodal distributions with skewnesses of 1.7 for fitting and 3.6 for validation.
We observe a small bias toward overestimating $h^{\mathrm{DFT}}$, with the constructed model predicting larger values for 57\% of samples in the  fitting set and 58\% of samples in the validation set.
The unimodal distribution of errors indicates that there is no subgroup of initial structures for which the model consistently fails.
Further, the Spearman correlation coefficients between the DFT-calculated enthalpies per atom and the model predicted enthalpies per atom are calculated as 0.88 for the fitting set and 0.87 for the validation set.

By considering the fractional error distributions in Figure\ \ref{fig:TriazineModelEnthalpy}(c) bias and non-Gaussian distribution in the constructed model is further shown.
The fitting set shows a mean fractional error of -0.016\,\% and the validation set has a mean fractional error of -0.019\,\%.
While the absolute errors show nearly symmetrical distributions, the fractional errors are skewed, with skewnesses of -3.7 for fitting and -8.6 for validation.
The skewness manifests as a tail in the fractional error distribution seen for $h^{\mathrm{DFT}} < h^{\mathrm{ML}}$.

The constructed random forest regressor models are able to predict volumes and enthalpies of relaxed structures of 1,3,5-triazine HCl over a range of number of constituent components in the unit cell.
MAE values, MAFE values and Spearman coefficients for the constructed models are similar between the fitting, validation, and testing sets.
These indicate that the models have not been overfit and therefore could be applied to unseen data.
The utility of the present machine learning approach comes from the ability to rank unrelaxed structures by the predicted relaxed volume or enthalpy then show further consideration only for the structures most likely to relax into high density, low enthalpy final configurations.
There is the potential to combine our machine learning property prediction with trained force-field approaches, see e.g. Refs \onlinecite{nikhar2022,metz2022,mattei2022}.
In such a hypothetical workflow, one would simultaneously train both the force-fields and the property prediction models using the same data.
Once both are trained, many trial structures would be generated and the property prediction model used as a filter to remove those which are likely to become thermodynamically unfavorable after optimization.

\begin{figure}
\includegraphics[width=\linewidth]{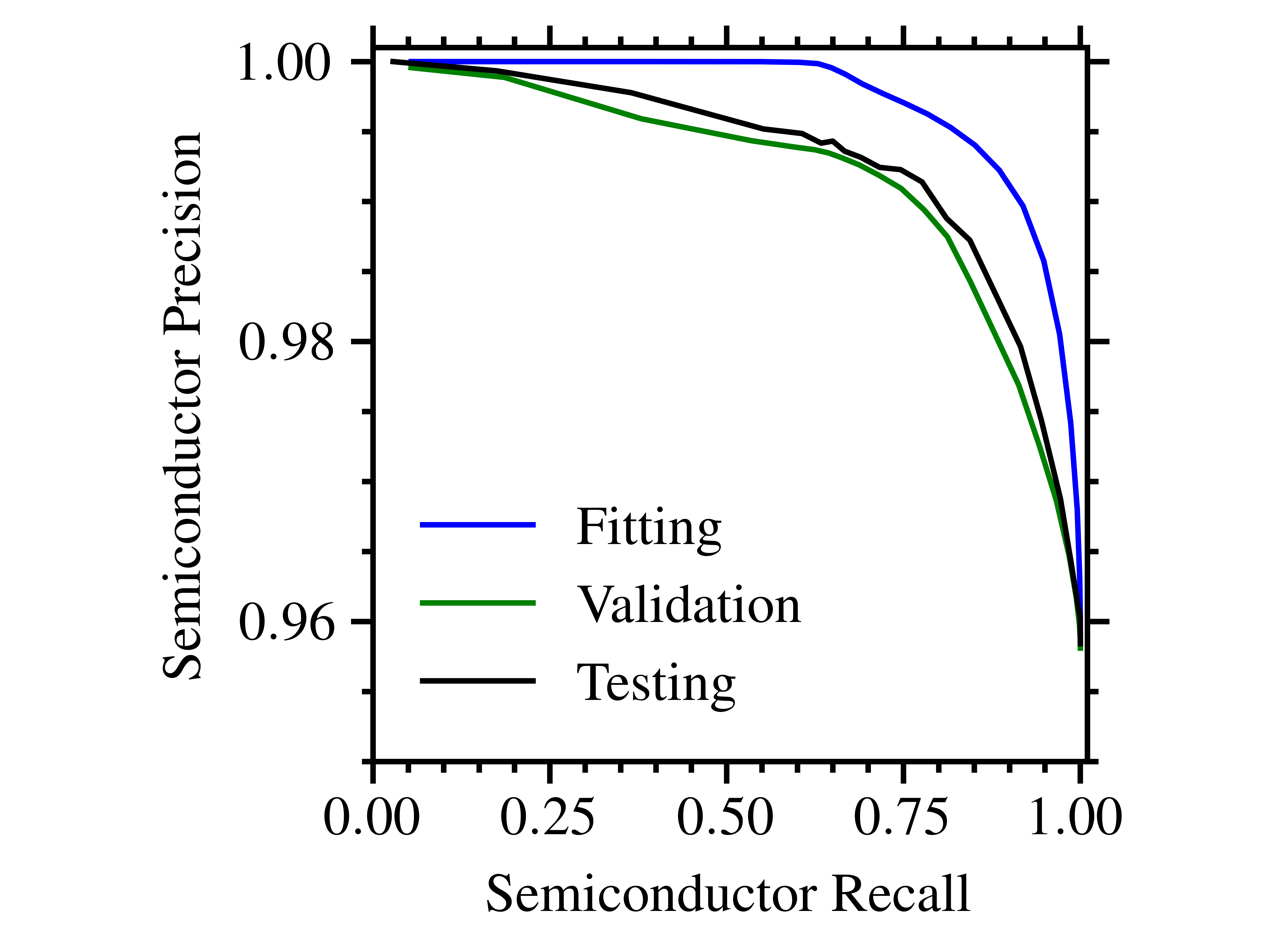}
\caption{\label{fig:TriazinePR}
  Precision-recall curve for prediction of semiconductor structures of 1,3,5-triazine HCl crystals.
}
\end{figure}

\textbf{Metal vs insulator model.} 
Material properties can be drastically and discontinuously altered by small changes which result in a metal-semiconductor transition\cite{gebhard1997}.
A material may have both metallic and semiconducting phases, corresponding to different crystal structures.
Based on the composition, it may be possible to anticipate which phase is more probable.
The phase model demonstrated here can be used to retain only structures in the more probable phase.
Removing structures likely to relax into the undesired metal or semiconductor phase can further speed up crystal structure prediction.
A grid search was performed over random forest classifiers with allowed maximum depths between 8 and 20 layers and minimum samples for splitting between 5 and 20 as classifiers for metallic versus semiconducting relaxed structures.
Using the mAP of the precision-recall curve as the optimization criterion, a maximum tree depth of 10 layers with a minimum splitting criterion of 10 samples was selected.
Figure \ref{fig:TriazinePR} plots the precision-recall curve for prediction of the semiconductor phase in crystal structures of 1,3,5-triazine HCl.
The model performs well in the case of the semiconductor phase with average precisions of 0.94 for the fitting and validation sets and 0.97 for the testing set.
The model struggles on the minority metal phase, with average precisions of 0.59 for the fitting set, 0.24 for the validation set, and 0.26 for the testing set.
These results give mAP values of 0.77 for the fitting set, 0.59 for the validation set, and 0.61 for the testing set.
The imbalances between the average precisions for semiconductors and metals indicates that the phase model prediction of a structure as a semiconductor is more reliable than the prediction as a metal.
This may be a consequence of the strongly imbalanced classes which could be improved by considering a chemical composition which has a more equitable division of semiconductor and metallic structures.

\subsection{\label{subsec:cgnn}Comparison with CGCNN}

\begin{figure}
\includegraphics[width=\linewidth]{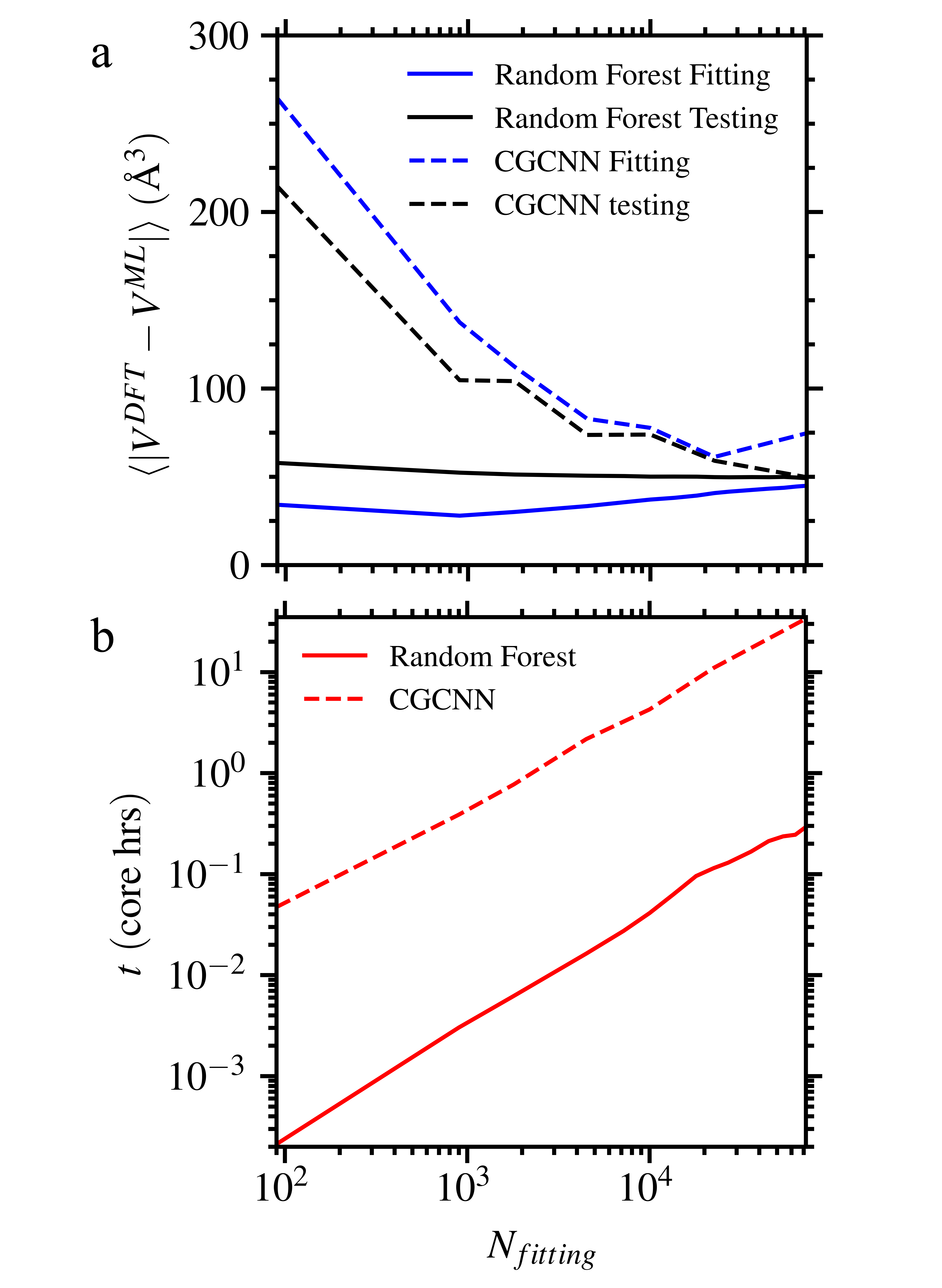}
\caption{\label{fig:AlgoPerformance}
  Comparison of performance of random forest regressor $V$ model and CGCNN $V$ model measured by (a) MAE and (b) fitting time.
}
\end{figure}

We benchmark the performance of the random forest regressor algorithm used here against the CGCNN approach by Xie \textit{et al.} as implemented in their publicly accessible code \cite{xie2018,xieGit}.
The two algorithms are compared by training models on the base model of 1,3,5-triazine HCl with varying numbers of structures in the fitting sets.
The results are plotted in Figure \ref{fig:AlgoPerformance}.
The models' fitting and testing MAEs are compared in Figure \ref{fig:AlgoPerformance} (a).
The random forest regressor has lower MAEs over the entire range of $N_{fitting}$ with the CGCNN only reaching the accuracy of the random forest regressor with over 70,000 structures in the fitting set.
Figure \ref{fig:AlgoPerformance} (b) plots the computational cost of fitting each model.
The time required to fit both models scales as $t\propto N_{fitting}^1$.
However, while the scaling with number of fitting samples is identical, the CGCNN algorithm requires two orders of magnitude more time to fit.
We also note that in the random forest approach 10 random forests are fit while the CGCNN approach fits only one neural network.
The random forest approach detailed in this work produces more accurate model with fewer structures needed for fitting and is fit significantly faster than the CGCNN.
These factors make the random forest approach preferred for scaling up to larger, chemically diverse training data.

\subsection{\label{subsec:extrapolate} Extending the Models}

In Section \ref{subsec:triazineModel}, three ML models predicting properties of 1,3,5-triazine HCl crystals were constructed and tested within an interpolative regime.
In order for a ML model to make reliable predictions for new chemical compositions, it must first see some samples with the new chemical composition.
Results in this subsection clarify the number of structures which are required to extend the use of the models and characterize their performance on the new chemical compositions.

\begin{figure}
\includegraphics[width=\linewidth]{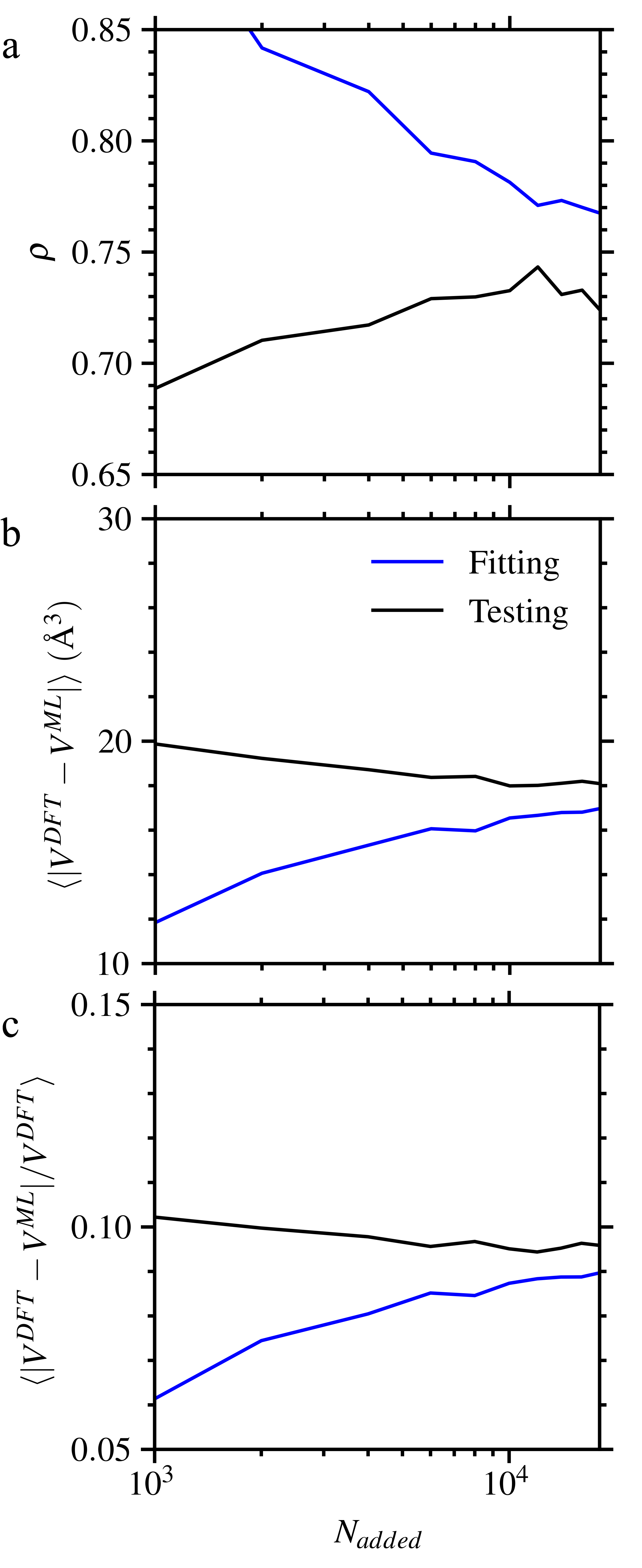}
\caption{\label{fig:123TriazineVolume}
(a) Spearman coefficient, (b) MAE, and (c) MAFE as 1,2,3-triazine HCl structures are added to the training set.
Values consider only structures of 1,2,3-triazine HCl.
}
\end{figure}

We examine a representative example of model performance as structures from a different chemical system are gradually added to a model. 
In Figure \ref{fig:123TriazineVolume}, structures of 1,2,3-triazine HCl are incorporated into the model training set.
The Spearman coefficient plotted in Figure \ref{fig:123TriazineVolume}(a) requires $\sim$ 10,000 added structures in order to converge.
With 10,000 structures of 1,2,3-triazine HCl added to the fitting set, the Spearman coefficient for the testing set is 0.73.
Adding up to 18,000 structures of 1,2,3-triazine HCl only increases the Spearman coefficient to 0.77.
The MAE and MAFE shown in Figures \ref{fig:123TriazineVolume}(b) and \ref{fig:123TriazineVolume}(c) converge with fewer added structures, requiring as few as 2,000 samples of 1,2,3-triazine HCl added to the fitting set.
These results demonstrate two important factors for training the models.
First, the number of structures from a new chemical composition which must be added to the base model for training is dependent on the measure used to evaluate the model.
Second, the models can be extended by adding as few as 2,000 to 10,000 structures with different chemical compositions.

Full results for extension tests are summarized in Tables \ref{table:AddToVolume}, \ref{table:AddToEnthalpy}, and \ref{table:AddToPhase}.
For both the volume and enthalpy per atom models, the MAFE values of the added structure testing sets are close to the base model MAFE values.
The effect of extension more significantly influences the values of the added structure testing set Spearman values.
The volume base model has a testing set Spearman coefficient of 0.95.
The testing set Spearman coefficients for the added structures decreases, to 0.72 - 0.73 for the A and B extension cases and to 0.62 and 0.40 for the C extension case.
Similarly for the enthalpy per atom models, the fitting MAE and MAFE values display limited variation between the base model and the models with added structures.
In the case of the enthalpy per atom models, the base model has a fitting set Spearman coefficient of 0.87.
Adding structures from new CSP runs approximately halves the fitting set Spearman coefficient for the added structures to 0.41 - 0.49 for the A and B extension cases and to 0.32 and 0.38 for the C extension case.
The opposite results are seen in the phase models.
The added structure fitting and testing set AUC values all increase compared to the base model fitting and testing AUC values.
We attribute this to the lower diversity of structures in the added CSP runs compared to the four CSP runs of 1,3,5-triazine HCl which contains several different combinations of organic molecule and HCl in the unit cell.

The model struggles with extending directly from the 1,3,5-triazine HCl to piperidine HCl.
The testing Spearman coefficient indicates a weak correlation between the values of $V^{DFT}$ and $V{ML}$.
The model is able to achieve a relatively low MAFE compared to other extensions by "guessing" the average value rather than learning a reliable relation between the initial structures and the final relaxed volume.
Future work will investigate if extension can be improved by training on a broader initial chemical space.

The decreases in testing set Spearman coefficients for the added structures fundamentally limits the accuracy of the ML approach.
The low Spearman coefficient causes the constructed models to have difficulty ranking the volumes and enthalpies per atom of new structures.
While the model approach cannot be used alone to identify experimentally obtainable structures, it can be used as a tool for downselecting structures for further computational study.

\begin{table*}[t]
\begin{tabular}{c c c c c c}
\hline
Added CSP set & Extension & Testing MAE ($\AA^3$)& Testing MAFE & Testing $\rho$ & $V^{DFT}$ range ($\AA^3$)\\ \hline
Base Model & None & 49 & 0.098 & 0.95 & 90.68 - 1,760.64 \\
1,3,5-triazine HBr & A & 18 & 0.092 & 0.72 & 100.09 - 418.24 \\
1,2,3-triazine HCl & A & 18 & 0.095 & 0.73 & 109.16 - 497.59 \\
1,2,4-triazine HCl & A & 19 & 0.100 & 0.72 & 88.94 - 492.22 \\
Pyridine HCl & B & 18 & 0.090 & 0.73 & 122.97 - 459.04 \\
Thiophene HCl & C & 18 & 0.095 & 0.62 & 107.46 - 505.52 \\
Piperidine HCl & C & 16 & 0.046 & 0.40 & 283.02 - 464.91 \\
\hline
\end{tabular}
\caption{\label{table:AddToVolume}
Performance of volume model with 10,000 structures from an extension CSP run added to the fitting set.
Each metric only measures the performance of the model on structures from the extension CSP run.
The "Base Model" row lists the results for the models described in Subsection \ref{subsec:triazineModel} for baseline comparison. 
}
\end{table*}

\begin{table*}[t]
\begin{tabular}{c c c c c c}
\hline
Added CSP set & Extension & Testing MAE (eV/atom) & Testing MAFE & Testing $\rho$ & $h^{DFT}$ range (eV/atom)\\ \hline
Base Model & None & 0.047 & 0.0076 & 0.87 & -7.01 - -3.20 \\
1,3,5-triazine HBr & A & 0.035 & 0.0057 & 0.44 & -6.39 - -4.92 \\
1,2,3-triazine HCl & A & 0.041 & 0.0068 & 0.43 & -6.31 - -4.00 \\
1,2,4-triazine HCl & A & 0.044 & 0.0072 & 0.41 & -6.39 - -3.71 \\
Pyridine HCl & B & 0.036 & 0.0060 & 0.49 & -6.28 - -4.90 \\
Thiophene HCl & C & 0.065 & 0.012 & 0.32 & -5.69 - -3.18 \\
Piperidine HCl & C & 0.015 & 0.0028 & 0.38 & -5.44 - -5.10 \\
\hline
\end{tabular}
\caption{\label{table:AddToEnthalpy}
Performance of enthalpy per atom model with 10,000 structures from an extension CSP run added to the fitting set.
Each metric only measures the performance of the model on structures from the extension CSP run.
The "Base Model" row lists the results for the models described in Subsection \ref{subsec:triazineModel} for baseline comparison.
}
\end{table*}

\begin{table*}[t]
\begin{tabular}{c c c c }
\hline
Added CSP set & Extension & Fitting mAP & Testing mAP \\ \hline
Base Model & None & 0.77 & 0.61 \\
1,3,5-triazine HBr & A & 0.74 & 0.41 \\
1,2,3-triazine HCl & A & 0.74 & 0.44 \\
1,2,4-triazine HCl & A & 0.77 & 0.40 \\
Pyridine HCl & B & 0.65 & 0.30 \\
Thiophene HCl & C & 0.66 & 0.48 \\
Piperidine HCl & C & 0.58 & 0.22\\
\hline
\end{tabular}
\caption{\label{table:AddToPhase}
Performance of phase model with 10,000 structures from an extension CSP run added to the fitting set.
Each metric only measures the performance of the model on structures from the extension CSP run.
The "Base Model" row lists the results for the models described in Subsection \ref{subsec:triazineModel} for baseline comparison.
}
\end{table*}

\section{\label{sec:conclusion}Conclusions}

In this work we have trained ML models to predict the properties of DFT-relaxed crystal structures of molecular salts based on only the unrelaxed structures.
The goal is to produce a machine learning method which filters out molecular crystal structures in CSP workflows by identifying which structures are likely to relax into physically unfavorable crystals.
We considered three key quantities: volume, enthalpy per atom, and metal versus semiconductor phase.
The chemical systems included small ring molecules of 1,2,3-triazine, 1,2,4-triazine, 1,3,5-triazine, pyridine, and thiophene mixed with varying concentrations of HCl.
Our approach has two key components to speed up model construction: we use crystal graph singular values instead of the full crystal graph representations, and random forests instead of neural networks.
Use of crystal graph singular values reduces the total number of descriptors by at least two orders of magnitude.
Random forests are fit more rapidly than neural networks and require tuning of fewer hyperparameters.
Each model is fit at low computational cost, each one requiring on the order of minutes to train on an individual workstation.
The structure evaluation and machine learning approach demonstrated in this work is not intended as a stand-alone CSP algorithm.
As presented, the ability to identify rare polymorphs would be slowed by the reliance on DFT for geometric optimization and randomly generated structures.
Instead, the utility of the machine learning approach is as a filtering step in other CSP efforts.
Integrating into other CSP efforts is beyond the scope of the presented work, but is the focus of ongoing study.

The models performed consistently well in the interpolative regime, with the testing and validation error distributions closely matching the fitting error distributions.
Performance of the models was inconsistent between target quantities in the extrapolative regime.
Testing volume and enthalpy per atom MAE and MAFE values for materials added to the base model were comparable to the testing MAE and MAFE values found for the base model.
Instead, difficulty in the extrapolative regime appeared as marked decreases in the Spearman coefficients between the DFT calculated and ML predicted values.
In the case of predicting semiconducting versus metallic phases, the models performed better on the added CSP runs.
Difficulty in extrapolating to new chemical spaces is typical of machine learning models.
Within our approach, this could be improved by broadening the chemical space included in the initial training set 

Our model building method shows several advantages compared to CGCNN approach.
While the time complexity for both neural networks and random forests is linear in the dimensionality of the material representation \cite{breiman2001,scikit-learn}, the computational cost of fitting the CGCNN is two orders of magnitude larger than fitting the random forest.
Further, the random forest regressors produce lower error models for smaller fitting sets.
Our set of crystal graph singular value descriptors accelerates model construction compared to the full crystal graph representation by reducing the required number of descriptors needed to characterize each material, while improving the accuracy of models fit with multiple chemical compositions.
While both neural networks (e.g. Refs.  \onlinecite{xie2018,ye2019,feng2019,kim2020}) and random forests (e.g. Refs. \onlinecite{nagasawa2018,takahashi2018,wang2020,goodall2020}) have shown success in predicting materials' properties, random forests tend to be easier to train due to requiring tuning of fewer hyperparameters.
Lastly, ML models can be constructed by training a single base model using a large number of structures in one chemical system, then taught to extrapolate to new chemical systems by incorporating data from approximately 2,000 to 10,000 structures.

A limitation of our chosen target quantities for CSP is that it assumes the experimentally obeservable polymorph can be determined from only thermodynamic considerations.
There are many cases among pharmaceutical molecules in which the thermodynamically most stable structure is kinetically hindered, and therefore not observed \cite{neumann2018}.
With a different choice of target quantities, one could also filter based on kinematic stability.
If these quantities are readily calculable, it would be possible to construct large datasets and employ our developed machine learning approach to include kinetic considerations as well.
A workflow could take the approach of identifying polymorph prototypes\cite{butler2023} then carrying out molecular dynamics simulations to determine the kinetically favored structure.

The model building approach taken in this work is general and can be extended in multiple directions.
A wider range of organic molecule components can be tested and incorporated into the models' training sets.
Target values and optimization criteria can be refined to better search for experimentally realizable polymorphs.
With the model training set sufficiently expanded, it can rank proposed polymorph structures to downselect which structures should receive further computational examination.
Our approach could be extended to include more complex systems: larger organic molecules, cocrystals, intercallated systems, and organometallic complexes.
Finally, the machine learning approach here is not limited to using quantities predicted with DFT.
It could instead be combined with data generated using, as an example, force-field methods\cite{neumann2008,nyman2016,mattei2022}.

\section*{Acknowledgments}
We thank Chris J. Pickard for assistance with \texttt{AIRSS}.
We thank Vinay Hegde and Bryce Meredig for helpful discussions.
Computational time was provided by the dCluster of the Graz University of Technology.
Funding was provided by Enterprise Science Fund, Intellectual Ventures.

\bibliography{./literature}

\subsection*{Contributions}
E.P.S. and C.H ran high throughput crystal structure prediction calculations.
E.P.S. developed the machine learning approach.
D.K.B. determined the chemical systems to study and design criteria.
E.P.S. and C.H. contributed in writing the manuscript.
All authors contributed to final editing of the manuscript.
R.P. and C.H. procured funding and supervised the project.

\subsection*{Competing Interests}
The authors declare no competing interests.

\subsection*{Data Availability}
POSCAR and .cif files of the initial unrelaxed structures used in this paper can be found in the Materials Data Facility\cite{MDF,blaiszik2019,shapera2023}.
Also included are datafiles containing relevant properties of the corresponding fully relaxed structures.

\subsection*{Code Availability}
Machine learning code is available from the corresponding author upon reasonable request.

\clearpage

\appendix
\renewcommand{\thesubsection}{S\arabic{subsection}}
\renewcommand{\theequation}{S\arabic{equation}}
\renewcommand{\thetable}{S\arabic{table}}
\renewcommand{\thefigure}{S\arabic{figure}}
\setcounter{equation}{0}
\setcounter{table}{0}
\setcounter{figure}{0}
\section*{Supplemental Materials}

\subsection{\label{subsec:MoleculeChoice} Choice of Molecular Crystals}

We focus on crystals formed from single ring heterocyclic organic molecules and HCl.
The specific molecules are listed in Table \ref{table:molecules} and shown in Figure\ \ref{fig:Molecules}.
1,3,5-triazine as the base constituent organic molecule is chosen because its small size lowers the computational cost of relaxing large number of crystals and allows for consideration of structures with multiple molecules in the unit cell. 
Triazine is a component of many larger organic molecules\cite{allen1996,groom2016} used in industry for manufacturing resins \cite{diem2010} and dyes \cite{tappe2000}, and as an organic reagent \cite{bohme2008}.
Further, there are two other molecules, 1,2,3-triazine and 1,2,4-triazine, which differ from 1,3,5-triazine by only the arrangement of N and C atoms in the ring.
We therefore use 1,2,3-triazine and 1,2,4-triazine as model compounds to test the ability of the model to extrapolate.

The ring of each organic molecule contains at least one heteroatom with a lone pair not part of the conjugated $\pi$-system, providing a site which may be protonated by HCl.
The organic molecule becomes protonated and forms an ionic bond with the Cl$^-$.
Controlling the concentration of HCl relative to the organic molecule leads to the formation of crystals with structures that may be manipulated through chemical means.

\begin{table*}[t]
\begin{tabular}{c c c c c}
\hline
Organic Molecule & Organic:HCl & Number of Structures & Fraction Metal & Always in Training? \\
\hline
1,3,5-triazine & 2:1 & 19,994 & 0.089 & Yes \\ 
1,3,5-triazine & 4:1 & 19,998 & 0.062 & Yes \\ 
1,3,5-triazine & 1:1 & 19,966 & 0.023 & Yes \\ 
1,3,5-triazine$^{*}$ & 3:3 & 29,991 & 0.011 & Yes \\ 
1,3,5-triazine & 1:1 (HBr) & 29,991 & 0.15 & No \\
1,2,3-triazine & 1:1 & 20,995 & 0.014 & No \\
1,2,4-triazine & 1:1 & 19,989 & 0.012 & No \\
Pyridine & 1:1 & 29,991 & 0.0068 & No \\
Thiophene & 1:1 & 49,940 & 0.020 & No \\
Piperidine & 2:2 & 28,325 & 0.0016 & No \\
\hline
\end{tabular}
\caption{\label{table:molecules}
Compositions of generated crystal structures.
$^{*}$Enforced cubic symmetry for starting structures.}
\end{table*}

\subsection{\label{subsec:CSP} Crystal Structure Prediction}

Random unrelaxed crystal structures are generated using the \texttt{AIRSS} \cite{pickard2006,pickard2011} software package.
The input structure contains a specified number of both protonated and unprotonated organic molecules along with the number of Cl$^-$ ions equal to the number of protonated organic molecules.
Structures are generated without imposing any symmetry constraints.
In the initial structures, organic molecules are generated with fixed geometries.
The relaxation process allows for optimization of the unit cell geometry, organic molecule shape, bonding, and relative positions and orientations of the molecules.

The generated crystal structures were relaxed with the \texttt{Vienna Ab-Initio software Package} (\texttt{VASP}) version 5.4.4 \cite{Kresse1996,Kresse1999}.
The exchange-correlation functional used is the generalized gradient approximation of Perdew, Burke, and Ernzerhof \cite{Perdew1992}. 
Corrections due to van der Waals forces are included using the DFT-D2 method of Grimme \cite{grimme2006}.
All DFT-calculations are carried out with a plane wave cutoff of 520 eV on $\Gamma$-centered k-meshes with a minimum separation between k-points of 0.5 $\AA^{-1}$.
Relaxing crystals of organic molecules poses a particularly difficult problem for DFT software.
The crystal unit cells may undergo significant changes in volume and shape, resulting in reciprocal lattice vectors computed in the initial stage of the relaxation being poor matches for the structure at the end of the relaxation.
To account for the notable discrepancies, relaxation calculations are carried out in a six-step iterative process with increasingly dense k-meshes and more allowed ionic steps.
The relaxed output structure at each step is used as the initial input structure for the subsequent relaxation step.
In the first five steps, structures are relaxed when either the change of the total energy is less than $2\times 10^{-3}$ eV or a fixed number of ionic steps has been reached.
In the sixth iteration, relaxation proceeds until the change of the total energy is less than $10^{-4}$ eV and the change in the total force is less than $10^{-3}$ eV/\AA.

The final relaxed structures from each CSP run are mixes of crystals consisting of discrete molecules, 1-D chains, and linked networks.
The diversity of final structures presents a worst-case scenario for fitting the models.
However, it would be expected to improve the ability of the model to generalize.

\subsection{\label{subsec:modelConstruct} Model Construction}

Fitting, validation, and testing errors were computed for tree-based ML algorithms, with and without the inclusion of boosting, available in \texttt{scikit-learn} \cite{scikit-learn}.
Among the considered ML algorithms, random forest regressors without boosting were found to produce the lowest error models with large positive Spearman correlation coefficients, defined below.
Large positive values indicate that the ML and DFT approaches produce similar orderings for each property.
In addition, this approach has the additional benefit of being readily optimized due to the small number of hyperparameters.
Constructed models were validated using a two-stage scheme, where 10\% of the training set was randomly designated the testing set.
The testing set remained fixed throughout validation and model construction.
The remaining training data was then divided randomly by 80\% into a set for fitting a decision tree regressor and 10\% into a validation set.
The fit model then was applied to both the validation and testing sets.
The process was repeated for 10 random fitting-validation partitions and values for the testing set were determined by averaging over these.

Models were constructed for three properties, cell volume ($V$), enthalpy per atom ($h$), and metal or semiconductor phase (phase).
These choices of target properties allow benchmarking of our approach for an extensive property, an intensive property, and a categorical property.
Metal versus semiconductor phase was determined for each structure by counting the number of electronic bands in the final relaxed structure below the Fermi level at each \textbf{k}-point.
Structures in which the number of bands below the Fermi level was constant across the Brillouin zone were labeled as 'semiconductor' while structures in which the number of bands below the Fermi level changes were labeled as 'metallic'.

\subsection{\label{subsec:modelEvaluate} Model Evaluation}

Once constructed, regression models were evaluated based on the mean absolute error (MAE), mean absolute fractional error (MAFE) \cite{sammut2011}, and Spearman correlation coefficient ($\rho$)\cite{coefficient2008} between the DFT-calculated values and the ML predicted values.
The MAE and MAFE are defined as:
\begin{equation}
\label{eq:MAE}
    MAE = \langle |t_i^{DFT}-t_i^{ML}| \rangle
\end{equation}
and
\begin{equation}
    \label{eq:MAFE}
    MAFE = \langle |\frac{t_i^{DFT}-t_i^{ML}}{t_i^{DFT}}| \rangle,
\end{equation}
where $t_i^{DFT}$ is the value of the target quantity for the i$^{th}$ material computed with DFT and $t_i^{ML}$ is the value predicted by the ML algorithm.
A consequence of the ML and DFT calculated values not matching is that the resulting orderings for each target value will also differ.
The extent to which two orderings match is characterized by the Spearman coefficient,
\begin{equation}
    \label{eq:spearman}
    \rho = 1 - \frac{6\sum_i \left[ R(t_i^{DFT}) - R(t_i^{ML}) \right]^2}{n(n^2-1)}.
\end{equation}
$R(x_i)$ is the rank function, which gives the position of $x_i$ in the ordered list of all $x$ values, and $n$ is the number of points in the list.
The Spearman coefficient takes values from +1 to -1, with +1 indicating the orders of the lists are identical and -1 indicating that the list orderings are reversed.

The reliability of a machine learning model to predict binary classification is often measured using the area under the Receiver-Operator Characteristic (ROC) curve \cite{bradley1997}.
In cases with strongly imbalanced classes, use of the precision-recall curve provides better sensitivity to the minority class \cite{goadrich2004,craven2004,kok2005,davis2006}.
The precision (P) and recall (R) of a binary classifier model for class $i$ are defined:
\begin{equation}
    P_i = \frac{TP}{TP + FP}
\end{equation}
and
\begin{equation}
    R_i = \frac{TP}{TP + FN}.
\end{equation}
$TP$ is the quantity of true positives identified by the model, $FP$ is the quantity of false positives, and $FN$ is the quantity of false negatives.
The average precision $AP$ for class $i$ is then
\begin{equation}
    AP_i = \int_{0}^{1}P_i(R_i) dR_i
\end{equation}
with a mean AP (mAP) characterizing the performance of the model for both classes:
\begin{equation}
    mAP = \frac{1}{2}(AP_1 + AP_2).
\end{equation}

\subsection{\label{subsec:includeCG} Inclusion of Crystal Graph Singular Value Descriptors}

\begin{figure}
\includegraphics[width=\linewidth]{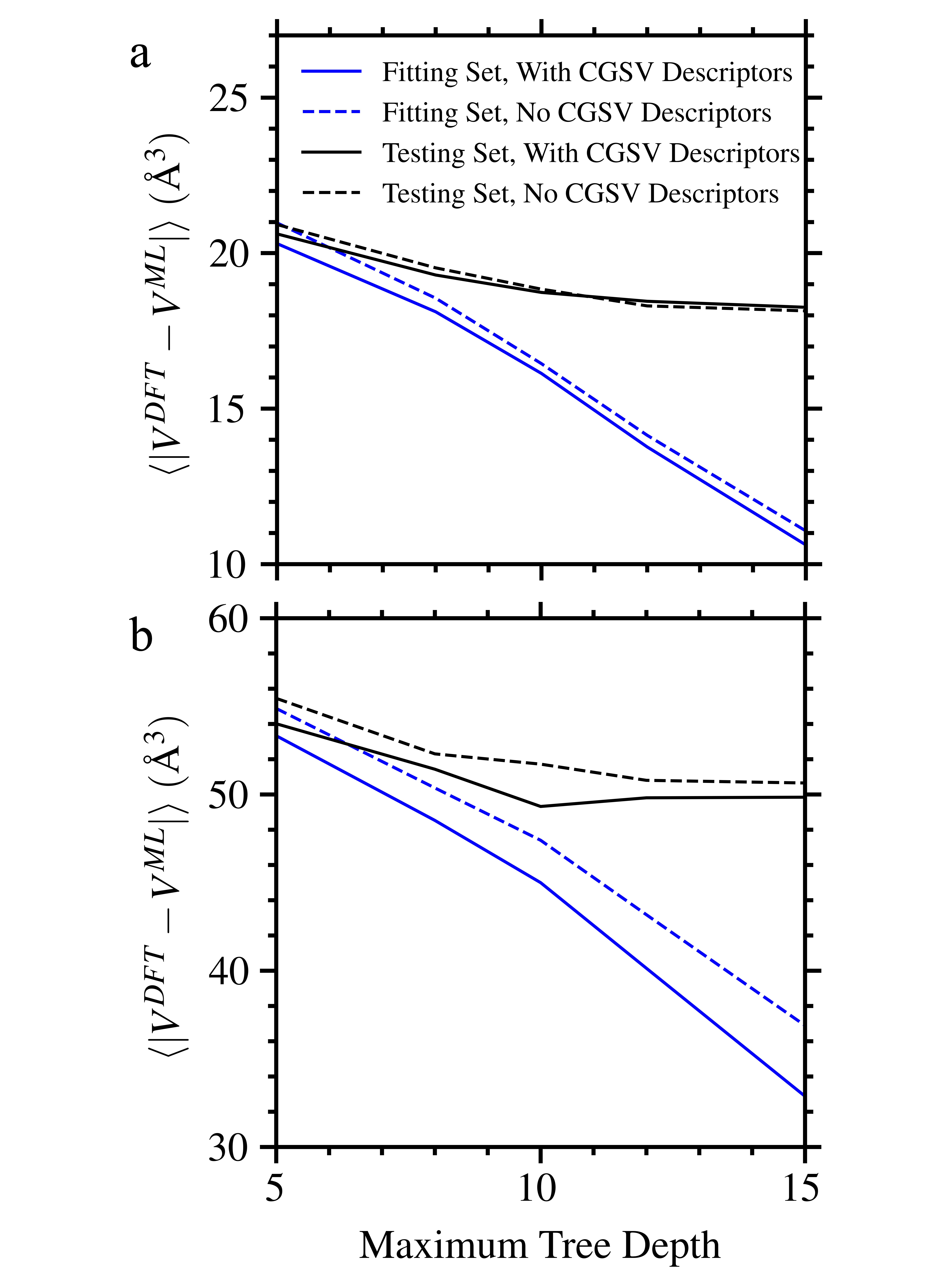}
\caption{\label{fig:descriptorchoice}
Comparison of MAEs for fitting and testing sets of $V$ models including (solid curves) and excluding (dashed curves) CGSV descriptors.
Model was trained on (a) structures of 1,3,5-triazine only and (b) the base model with four different unit cell compositions of 1,3,5-triazine HCl.
}
\end{figure}

Figure \ref{fig:descriptorchoice} compares the performances of volume models which include or exclude the CGSV descriptors.
In both cases models include Coulomb matrix and crystal structure descriptors and filter out low importance descriptors as described in Section \ref{subsec:descriptorSelect}.
In the 1,3,5-triazine model, shown in Figure \ref{fig:descriptorchoice}(a), inclusion of the CGSV descriptors causes a minimal change in the MAEs of the volume model.
For instance, at a maximum tree depth of 10 layers, without CGSV descriptors the model produces MAEs of 16.4 $\AA^3$ for the fitting set and 18.8 $\AA^3$ for the testing set.
Inclusion of CGSV descriptors lowers the MAEs to 16.1 $\AA^3$ and 18.7 $\AA^3$ for the fitting and testing sets, respectively.
In the case of the base model shown in Figure \ \ref{fig:descriptorchoice}(b), the volume model without CGSV descriptors is found to have MAEs of 47.4 $\AA^3$ for the fitting set and 51.7 $\AA^3$ for the testing set.
Inclusion of CGSV descriptors lowers the MAEs to 45.0 $\AA^3$ for the fitting set and 49.3 $\AA^3$ for the testing set.
The 1,3,5-triazine model includes a limited region of chemical space, only considering structures with four molecules of 1,3,5-triazine in the unit cell.
The base model includes four different unit cell compositions with three ratios of 1,3,5-triazine molecules to HCl molecules.
Inclusion of descriptors beyond the Coulomb matrix and crystal structure improve the accuracy of models which include multiple chemical compositions.

\subsection{\label{subsec:baseline} Baseline Test of CSP and Machine Learning}

\begin{figure}
\includegraphics[width=\linewidth]{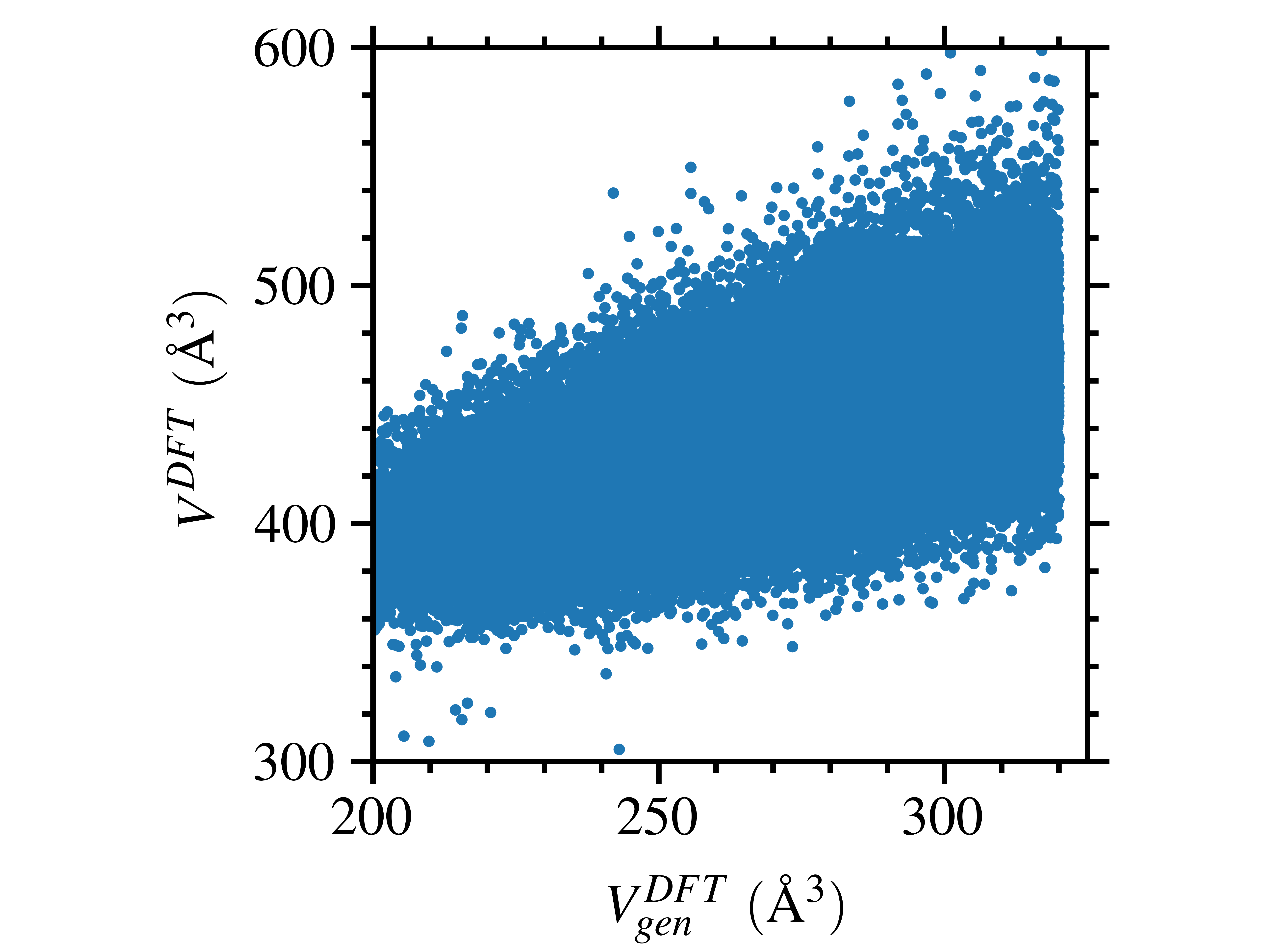}
\caption{\label{fig:initial-finalvolumes}
 Comparison of DFT relaxed volumes ($V^{DFT}$) and unrelaxed volumes ($V^{DFT}_{gen}$) of 60,000 \texttt{AIRSS}-generated structures.
 Cells consist of four molecules of 1,3,5-triazine.
}
\end{figure}

\begin{figure}
\includegraphics[width=\linewidth]{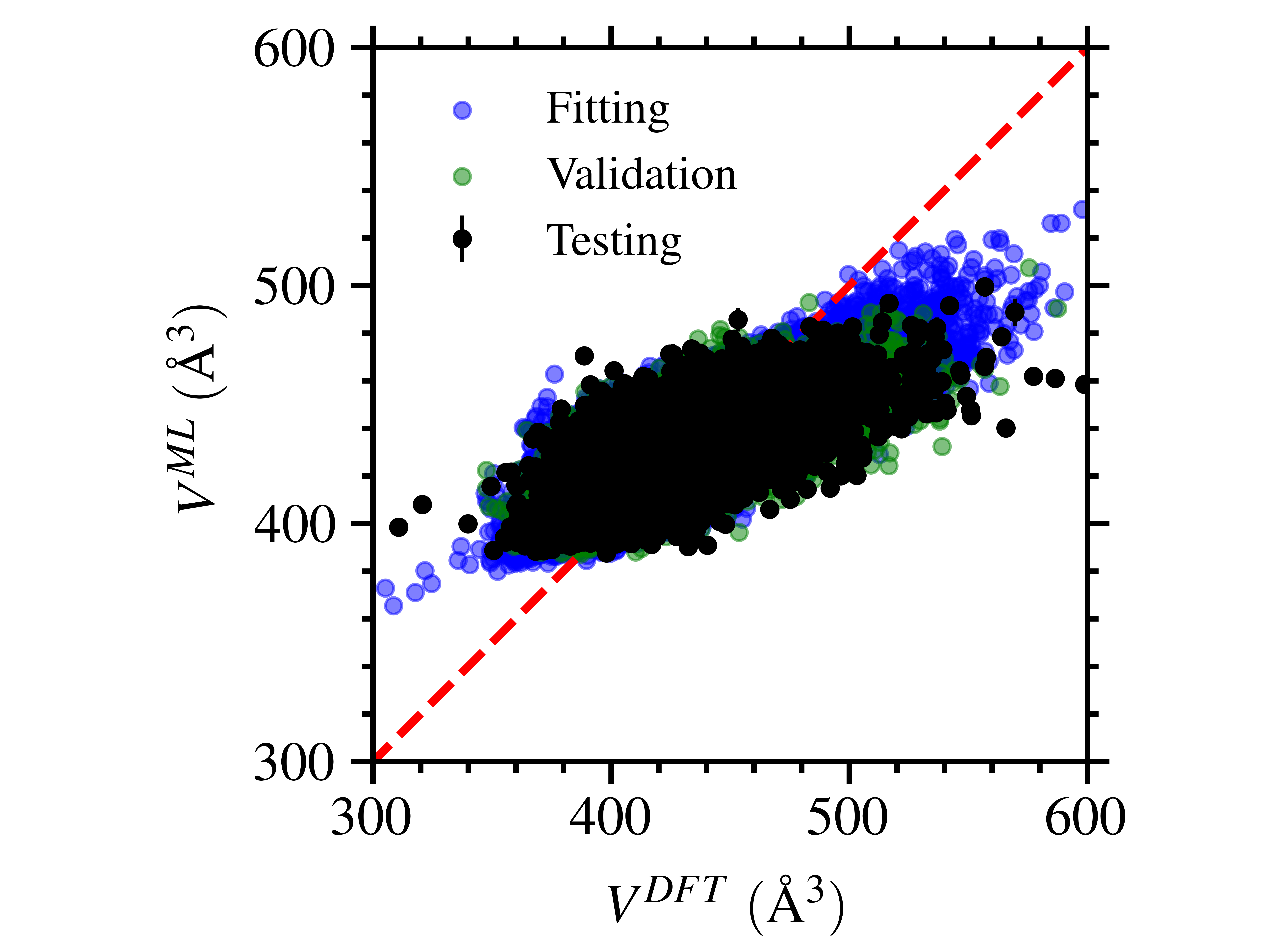}
\caption{\label{fig:triazineonlyVolumePlot}
 Parity plot comparison of $V^{DFT}$ obtained via DFT and  $V^{ML}$ predicted by the random forest regressor for the dataset of only 1,3,5-triazine.
}
\end{figure}

\begin{figure}
\includegraphics[width=\linewidth]{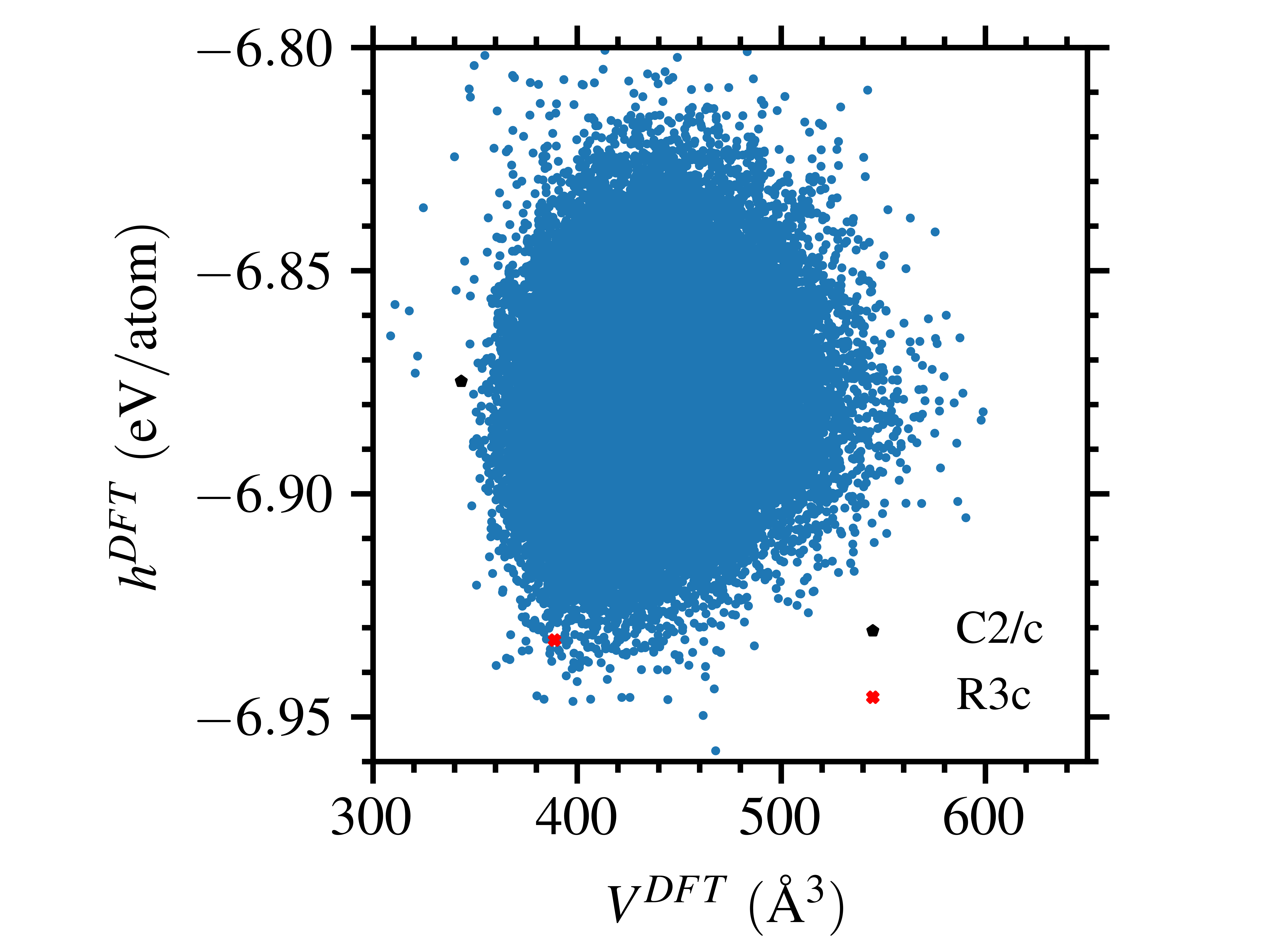}
\caption{\label{fig:triazineCSPvsExp}
 Distribution of $V^{DFT}$ and $h^{DFT}$ for 60,000 \texttt{AIRSS}-generated structures of 1,3,5-triazine after relaxation.
 The black pentagon marks the position of the experimental C2/c structure and the red x marks the R3c structure, with volume scaled to match the four molecule per cell configuration used for structure generation.
}
\end{figure}

As a baseline test of our CSP and machine learning approaches, we use the process for generating and relaxing structures described in Section \ref{subsec:CSP} to form a training dataset of crystals of 1,3,5-triazine.
Only structures with four molecules of 1,3,5-triazine are considered in this test.
The correlation between the unrelaxed and relaxed volumes is 0.59.
The machine learning model for $V$ produces Spearman coefficients of 0.78 for the fitting set, 0.67 for the validation set, and 0.66 for the testing set.
The trained model outperforms simply selecting the lowest volume generated structures.

The choice of 1,3,5-triazine without added acid allows benchmarking of the relaxation process against known experimental crystal structures of 1,3,5-triazine.
At room temperature the experimentally observed polymorph has an R3c structures.
With increasing pressure, crystals of 1,3,5-triazine undergo two structural phase transitions with: first from an R3c structure to a C2/c structure near 0.6 GPa \cite{eckert1982,dove1985,citroni2014}, then to an amorphous structure at 15.2 GPa \cite{li2014}.
Figure \ref{fig:triazineCSPvsExp} shows the distribution of $h^{DFT}$ versus $V^{DFT}$ for 60,000 structures of 1,3,5-triazine after relaxation.
The R3c and C2/c structures of 1,3,5-triazine have been taken from the Cambridge Structural Database\cite{groom2016} and relaxed using the same settings as the final step of our relaxation process.
The low pressure R3c structure lies in the low enthalpy, low volume region of the diagram, which is consistent with the R3c structure being realizable.

\subsection{\label{subsec:global} Global versus Local Models}

\begin{figure}
\includegraphics[width=\linewidth]{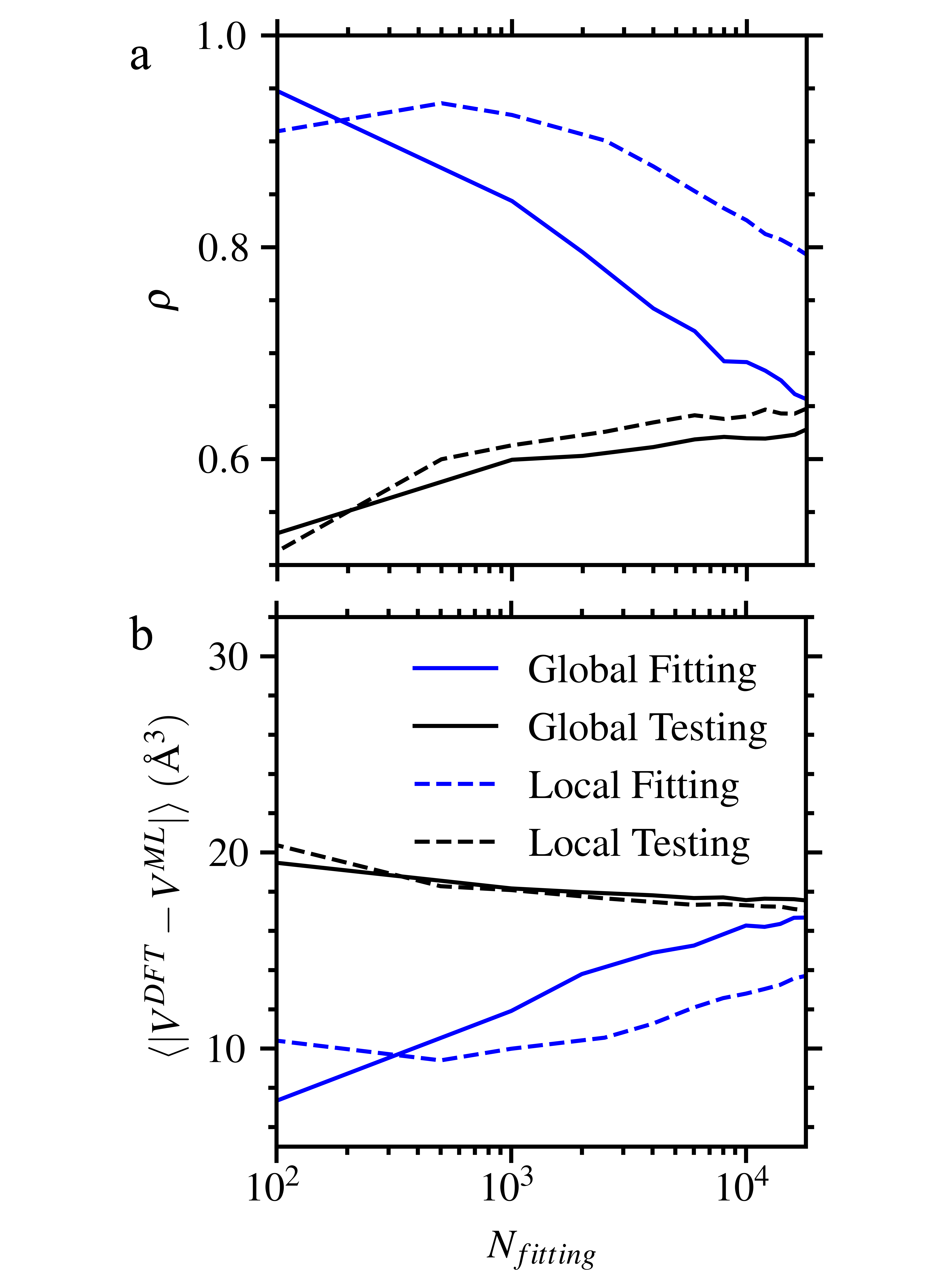}
\caption{\label{fig:globalvlocal}
 Comparison between global and local $V$ models for thiophene HCl.
 Solid curves correspond to models trained using the base model with thiophene HCl structures incrementally added.
 Dashed curves correspond to models trained with only thiophene HCl
 Values consider only the thiophene HCl structures.
 The models are compared using (a) the Spearman coefficient and (b) the MAE.
}
\end{figure}

The performance of a 'global' machine learning model is compared against a 'local' model trained on the same dataset.
The 'global' model was trained on the 'base' crystal structure set of four different compositions of 1,3,5-triazine and HCl and structures of thiophene HCl incrementally added to the fitting data.
The 'local' model was trained with varying numbers of structures of thiophene HCl in the fitting set.
Figure \ref{fig:globalvlocal} (a) compares the MAEs and Figure \ref{fig:globalvlocal} (b) compares the Spearman coefficients of the 'global' and 'local' models as the number of thiophene HCl structures in each fitting set is increased.
Above $N=1,000$, both the MAE and $\rho$ indicate that the 'local' model outperforms the 'global' model on the fitting set.
For instance, at $N=10,000$, the 'local' model fitting set has an MAE of 12.8 $\AA^3$ and Spearman coefficient of 0.83.
This compares against the 'global' model fitting set MAE of 16.3 $\AA^3$ and Spearman coefficient of 0.69.
The two models perform nearly equally well for the testing sets.
The 'local' model testing set has an MAE of 17.3 $\AA^3$ and Spearman coefficient of 0.64 while the 'global' model testing set has an MAE of 17.6 $\AA^3$ and Spearman coefficient of 0.62.
We also note that the MAEs and Spearman coefficients for the base data set in the 'global' model is identical within significant figures to the values found in the model trained using only the base dataset.

The 'local' model outperforms the 'global' model at fitting the data due to the smaller dataset size and lower chemical diversity used to train the 'local' model.
However, both models show nearly identical performance when applied to unseen data in the testing sets.
This shows the potential to construct global machine learning models over broader chemical spaces with a minimal loss of reliability compared to training individual models over restricted chemical spaces.
This broadening of the training dataset does not decrease the predictive power of the model for structures included in the base of the 'global' model.
It may be possible to train a single 'global' model on a sufficiently broad chemical space that it could extrapolate to new chemical compositions.

\end{document}